\pdfoutput=1 
\documentclass{aa}  
\usepackage{graphicx}
\usepackage[varg]{txfonts}
\usepackage{natbib}
\usepackage[normalem]{ulem}
\begin{document}

\def\ltsima{$\;\buildrel < \over \sim \;$}
\def\simlt{\lower.5ex \hbox{\ltsima}}
\def\gtsima{$\;\buildrel > \over \sim \;$}
\def\simgt{\lower.5ex \hbox{\gtsima}}
\newcommand{\new}[1]{#1}
\newcommand{\vn}[1]{#1}
\newcommand{\st}[1]{}

   \title{Ubiquitous argonium (ArH$^+$) in the diffuse interstellar medium \\-- a molecular tracer of almost purely atomic gas}

   \author{P.~Schilke
          \inst{1}
          \and
          D.A.\ Neufeld\inst{2}
          \and
          H.S.P.~M\"uller\inst{1}
          \and
          C.~Comito\inst{1}
          \and
          E.A.~Bergin\inst{3}
          \and
          D.C.~Lis\inst{4,5}
          \and 
          M.~Gerin\inst{6}
          \and
          J.H.~Black\inst{7}
          \and
          M.~Wolfire\inst{8}
          \and
          N. Indriolo\inst{2}
          \and
          J.C.~Pearson\inst{9}
          \and 
          K.M.~Menten\inst{10}
          \and
          B.~Winkel\inst{10}
          \and
          \'A. S\'anchez-Monge\inst{1}
	  \and
	  T. M\"oller\inst{1}
	  \and
	  B.~Godard\inst{6}
	  \and
	  E.~Falgarone\inst{6}
          }

   \institute{I.~Physikalisches Institut der Universit\"at zu K\"oln, Z\"ulpicher Str.~77, 50937 K\"oln, Germany\\
              \email{schilke@ph1.uni-koeln.de}
         \and
             The Johns Hopkins University, Baltimore, MD 21218, USA
             \and
             Department of Astronomy, The University of Michigan, 500 Church Street, Ann Arbor, MI 48109-1042, USA
             \and
             California Institute of Technology, Pasadena, CA 91125, USA
    \and 
	    Sorbonne Universit\'{e}s, Universit\'{e} Pierre et Marie Curie, Paris 6, CNRS, Observatoire de Paris, UMR 8112, LERMA, Paris, France
             \and
             LERMA, CNRS UMR8112, Observatoire de Paris \& Ecole Normale Sup\'erieure, 24 rue Lhomond, 75005, Paris, France
             \and
             Department of Earth and Space Sciences, Chalmers University of Technology, Onsala Space Observatory, 439 92 Onsala, Sweden
             \and
             Astronomy Department, University of Maryland, College Park, MD 20742 USA 
	    \and
	     Jet Propulsion Laboratory, California Institute of Technology, Pasadena, CA 91109, USA
             \and
             Max-Planck-Institut f\"ur Radioastronomie, Auf dem H\"ugel 69, D-53121 Bonn, Germany
             }

   \date{}

 
  \abstract
   {}
{We describe the assignment of a previously unidentified interstellar absorption line to ArH$^+$ and discuss its relevance in the context of hydride absorption in diffuse gas with a low H$_2$ fraction.
   The confidence of the assignment to ArH$^+$ is discussed, and the column densities are determined toward \new{several} lines of sight.  The results are then discussed in the framework of chemical models, with the aim of explaining the observed column densities.}
   {We fitted the spectral lines  with multiple velocity components, and determined column densities  from the line-to-continuum ratio.  The column densities of ArH$^+$ we compared to those of other species, tracing \new{interstellar medium (ISM)} components with different H$_2$ abundances. We constructed chemical models that take UV radiation and cosmic ray ionization into account.}
   {Thanks to the detection of two isotopologues, $^{36}$ArH$^+$ and $^{38}$ArH$^+$, we are confident about the carrier assignment to ArH$^+$. NeH$^+$ is not detected with a limit of [NeH$^+$]/[ArH$^+$] $\le$ 0.1. The derived column densities agree well with the predictions of chemical models. ArH$^+$ is a unique tracer of gas with a fractional H$_2$ abundance of $10^{-4}- 10^{-3}$ and shows little correlation to H$_2$O$^+$, which traces gas with a fractional H$_2$ abundance of $\approx $0.1.}
   {A careful analysis of variations in the ArH$^+$, OH$^+$, H$_2$O$^+$, and HF column densities promises to \new{be a faithful tracer} of the distribution of the H$_2$ fractional abundance by providing unique\new{ information on} a poorly known phase in the cycle of interstellar matter and on its transition  from atomic diffuse gas to dense molecular gas traced by CO emission. Abundances of these species put strong observational constraints upon \new{magnetohydrodynamical (MHD) } simulations of the interstellar medium, and potentially could evolve into a tool characterizing the ISM. Paradoxically, the ArH$^+$ molecule is a better tracer of \new{almost} purely atomic hydrogen gas than H{\sc i} itself, since \new{H{\sc i} can also be present in gas with a significant} molecular content, but ArH$^+$ singles out gas that is $>99.9$\% atomic. }

               
  \titlerunning{Ubiquitous argonium (ArH$^+$) in the diffuse interstellar medium}
  \authorrunning{Schilke et al.}

   \maketitle
%

\section{Introduction}

Light hydrides of the type ZH$_{\rm n}$ or ZH$_{\rm n} ^+$ are important diagnostics of 
the chemical and physical conditions in space. Their lower energy rotational transitions 
occur for the most part at terahertz frequencies (far-infrared wavelengths). 
This frequency region can be only accessed  to a limited extent from the ground, even at 
elevated sites, because of strong atmospheric absorptions of H$_2$O and, to a lesser 
degree, O$_2$ and other molecules. The \textit{Herschel} Space 
Observatory \citep{Herschel_2010} has provided a powerful new probe of the submillimeter and far-infrared spectral regions, which greatly expands upon the capabilities afforded by \new{earlier} missions, such as the Kuiper Airborne Observatory 
\citep[KAO;][]{KAO_1976}, the Infrared Space Observatory \citep[ISO;][]{ISO_1996}, and others, 
or from ground with the Caltech Submillimeter Observatory \citep[CSO;][]{CSO} or the Atacama Pathfinder 
EXperiment \citep[APEX;][]{APEX_2006}.

Observations of hydride molecules, in particular of H$_2$O, in interstellar space, 
but also in solar system objects, were among the important goals of the \textit{Herschel} 
mission. In fact, the cationic hydrides H$_2$O$^+$ \citep{H2O+_det_2010}, H$_2$Cl$^+$ 
\citep{H2Cl+_det_2010}, and HCl$^+$ \citep{HCl+_det_2012} were detected with \textit{Herschel} 
for the first time in the ISM. While the SH radical has its fundamental transition at a frequency that was inaccessible to the high-resolution Heterodyne Instrument for Far-Infrared Astronomy 
\citep[HIFI;][]{HIFI_1_2010}, it has been detected \citep{SH_det_2012} with the German REceiver At 
Terahertz frequencies \citep[GREAT;][]{GREAT_2012} onboard the Stratospheric Observatory 
For Infrared Astronomy  \citep[SOFIA;][]{SOFIA-1_2012,SOFIA-2_2013}. 
OH$^+$ \citep{OH+_det_2010}. and SH$^+$ \citep{SH+_det_2011} were detected with APEX 
from the ground shortly before \textit{Herschel}, but many additional observations 
were carried out with HIFI, \citep[e.g.,][]{Godard2012}. Several hydrides, e.g., OH$^+$ and H$_2$O$^+$, were found 
to be widespread with surprisingly high column densities, not only in Galactic sources 
\citep{OH_n^+_2010,H2O+_det_2010,Neufeld2010}, but also in external galaxies
\citep{det_H2O+_M82_2010,det_OH+_Mrk231_2010,OH_n^+_Arp220_2013}. As both cations react fast 
with H$_2$ to form H$_2$O$^+$ and H$_3$O$^+$, respectively, it was suspected that these molecules 
reside in mostly atomic gas, which contains little H$_2$ \citep{OH_n^+_2010}. Detailed model calculations 
suggest that the abundances of OH$^+$ and H$_2$O$^+$ are particularly high in gas with 
molecular fraction of around 0.05 to 0.1 \citep{Neufeld2010,OH_n+_chemistry_2012}. 
The comparatively high column densities observed for these two molecular cations also 
require cosmic ray ionization rates in the diffuse ISM to be considerably higher than 
that in the dense ISM \citep{Neufeld2010,OH_n+_chemistry_2012,CRI-rate_W51_2012}.
However, \new{evidence for high ionization rates in the diffuse ISM, in the range $10^{-16} - 10^{-15}$ s$^{-1}$, has been presented already earlier to explain the amount of 
H$_3 ^+$ in the diffuse ISM \citep{H3+_higher-CRI-rate_1998,models_CRI-rate_diff_2003, IndrioloMcCall2012} }.
Even higher cosmic ray ionization rates were estimated for active galaxies such as 
NGC~4418 and Arp~220 \citep[][$> 10^{-13}$ s$^{-1}$]{OH_n^+_Arp220_2013}.

Spectral line surveys of the massive and very luminous Galactic Center sources Sagittarius~B2(M) 
and (N)  were carried out across the entire 
frequency range of HIFI within the guaranteed time key project HEXOS, \citep{HEXOS_2010}. A moderately strong absorption feature was detected toward both sources 
near 617.5~GHz, but the carrier proved very difficult to assign \citep{Schilke2010, iaudib}. This feature appears at all velocity components associated with diffuse, foreground gas, but is conspicuously absent at velocities related to the sources themselves, suggesting that the carrier resides only in very diffuse gas.
The absorption line was detected toward other continuum sources as well
during subsequent dedicated observations (within the guaranteed time key project PRISMAS; \citealp{OH_n^+_2010, iaudib}).

Very recently, \citet{Barlow2013} observed a line in emission at the same frequency toward 
the Crab Nebula supernova remnant, which they assigned to the $J = 1 - 0$ transition of argonium $^{36}$ArH$^+$ at 617.525 GHz. 
 In addition, they observed the $J = 2 - 1$ transition at 1234.602~GHz. Here we present 
evidence that $^{36}$ArH$^+$ is also responsible for the absorption features detected in the HEXOS and PRISMAS spectra.

\section{Observations}
The 617.5~GHz features were first discovered in absorption in the full spectral
scans of Sagittarius B2(M)\footnote{Herschel OBSIDs: 1342191565} and (N)\footnote{Herschel OBSIDs: 1342206364}  carried out between 2010 March and 2011 April
using \textit{Herschel}/HIFI, within the framework of the
HEXOS guaranteed time key Program. The data presented here have
been re-reduced using an improved version of the HIFI pipeline, which results in significantly
lower noise levels in the high-frequency HEB mixer bands. 
The line survey data have been calibrated with HIPE version 10.0 \citep{HIFI_2_2012} and the resulting double-sideband (DSB) spectra were subsequently reduced using
the GILDAS CLASS\footnote{http://www.iram.fr/IRAMFR/GILDAS} package. Basic data reduction steps included removal of spurious
features or otherwise unusable portions of the spectra. The continuum emission was then
subtracted from the DSB scans by fitting a low-order polynomial (typically first, in a few cases second order). The continuum-subtracted DSB data were deconvolved (sideband separation
through pure $\chi^2$ minimization; \citealp{Comito2002}) to provide a single-sideband (SSB)
spectrum for each HIFI band.
A linear least squares fit of the subtracted continuum values as a function of the
LO frequency provided a reliable (unaffected by spectral features) parametrization of
the continuum variability across each HIFI band, which was then folded back into the
deconvolved, continuum-subtracted SSB spectra.
Finally, the overall Sagittarius B2(M) and (N) continuum was rendered self-consistent in two
steps: the first adjustment consisted of an additive factor for each band, to achieve a
zero-continuum level for the saturated absorption features; the second adjustment required
a multiplicative factor, in order for the continuum values in overlap regions between bands
to be consistent with one another.

\section{Spectroscopy}

The noble gas hydride cations NgH$^+$, with Ng heavier than helium, are 
isoelectronic with the hydrogen halides HX; HeH$^+$ is isoelectronic with
H$_2$ with $^1 \Sigma ^+$ ground electronic states. All non-radioactive 
noble gas hydride cations have been thoroughly characterized both spectroscopically and kinetically. 


On Earth, $^{40}$Ar with an isotopic abundance of 99.6\,\% is by far the dominant isotope
\citep{composition_elements_2009}, but the terrestrial $^{40}$Ar 
originates almost exclusively from the radioactive decay of $^{40}$K. Solar and interstellar argon 
is dominated by $^{36}$Ar ($\sim$84.6\,\%), followed by $^{38}$Ar 
($\sim$15.4\,\%), with only traces of $^{40}$Ar ($\sim$0.025\,\%)
\citep{solar_noble_gases_2002}.

\subsection{Rest frequencies}
\label{Rest frequencies}

Rest frequencies of $^{36}$ArH$^+$, $^{38}$ArH$^+$, and $^{20}$NeH$^+$ (which was in the frequency range of the survey) were taken from 
the Cologne Database for Molecular Spectroscopy \citep[CDMS][]{CDMS_1,CDMS_2}\footnote{http://www.astro.uni-koeln.de/cdms/}. Extensive rotational and rovibrational data were 
critically evaluated and combined in one global fit for either molecular 
cation taking the breakdown of the Born-Oppenheimer approximation into account. 

The most important spectroscopic data in the case of ArH$^+$ are the measurements of 
rotational transitions of $^{40}$ArH$^+$ \citep{ArH+_rot_high-J_1988,ions_FIR_1987} 
and of the $J = 1 - 0$ transitions of $^{36}$ArD$^+$, $^{38}$ArD$^+$, and 
$^{40}$ArD$^+$ \citep{isos-ArD+_1-0_rot_1983}. Additional data comprise further 
rotational transition frequencies of $^{40}$ArD$^+$ \citep{ArD+_rot_1999}, 
as well as rovibrational data of $^{40}$ArH$^+$ \citep{ArH+_IR_1982,NgH-D+_IR_1984}, 
 $^{36}$ArH$^+$, and $^{38}$ArH$^+$ \citep{36-38ArH+_IR_1988}, and of 
$^{40}$ArD$^+$ \citep{NgH-D+_IR_1984}. Very recently, \citet{Cueto2014} reported rovibrational transition frequencies, which were rather accurate by infrared standards ($\approx 3-4$~MHz), but only of modest accuracy by microwave standards.  Inclusion of these data, therefore, did not change the frequencies and uncertainties significantly.

Rotational spectra of $^{20}$NeH$^+$, $^{22}$NeH$^+$, $^{20}$NeD$^+$, and $^{22}$NeD$^+$ were published by 
\citet{NeH+_rot_1998}. Additional, mostly rovibrational, data were taken from 
\citet{NeH+_IR_1982,NeH+_IR_1985,ions_FIR_1987,NeH+_IR_2004}. 
The electric dipole moments of ArH$^+$ and NeH$^+$ (2.2 D and 3.0 D, respectively) were taken from quantum chemical 
calculations \citep{hydride_cations_ai_2007}. Other transition frequencies used in the analysis were taken from the CDMS and JPL \citep{CDMS_1, CDMS_2, JPL} catalogs.  Specifically, the methanimine (H$_2$CNH) entry, based on \citet{Dore2012}, was taken from the CDMS, while the methylamine (CH$_3$NH$_2$) entry, based on \citet{Ilyushin2005}, was taken from the JPL catalog.

\section{Results}

We report here the detection of an absorption  line we identify with $^{36}$ArH$^+$(1--0) toward a number of strong 
continuum sources, viz.\ SgrB2(M) and SgrB2(N) from the HEXOS key Program \citep{HEXOS_2010}, G34.26+0.15, W31C (G10.62-0.39), W49(N), and W51e, from the PRISMAS key Program \citep{OH_n^+_2010}. These sources are star-forming regions that provide strong continuum background illumination for absorption studies of foreground material.  \new{Their being star-forming regions}, however opens also the possibility of the background continuum being contaminated by source-intrinsic emission lines. The molecular cation is observed at all velocities 
corresponding to diffuse molecular clouds on the line of sight toward 
these sources, but is conspicuously absent \new{(or very weak) }at velocities related to the 
sources themselves. We report also upper limits 
on the column densities of $^{20}$NeH$^+$ toward SgrB2(M) and (N).

\begin{figure*}
\centering
\includegraphics[width=18cm]{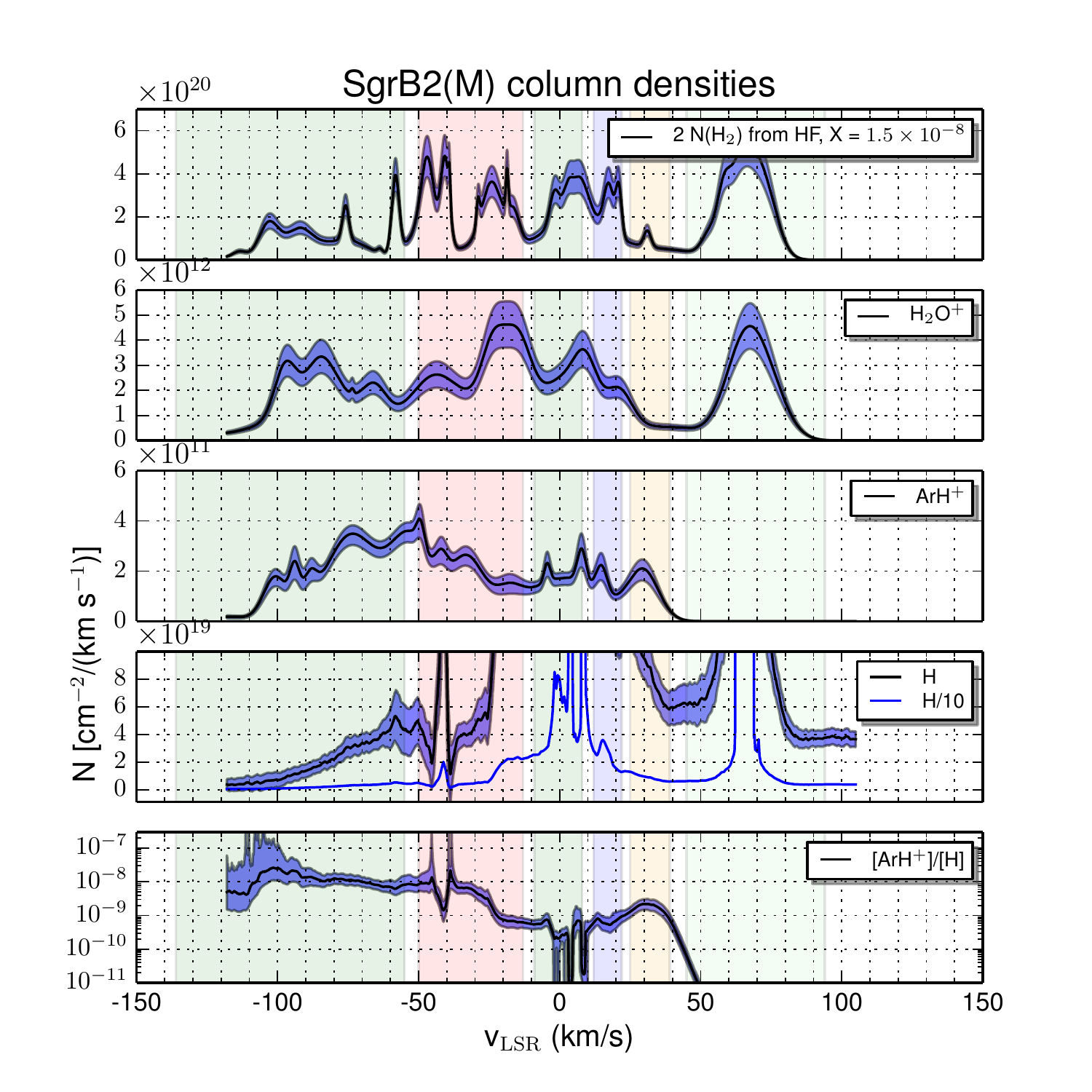}
\caption{Column density per km s$^{-1}$ of HF, H$_2$O$^+$, ArH$^+$, and H, in descending order of f(H$_2$) traced by the species.  \new{The color coding of the frequencies is explained in Tab.~\ref{tab:distances}. The error estimate for ArH$^+$ was done using the MAGIX Interval Nested Sampling algorithm \citep{Moeller2013}, which implements a Markov chain Monte Carlo (MCMC) method to calculate the Bayesian evidence and Bayesian confidence interval. H{\sc i} column density errors were calculated by propagating uncertainties
from the input emission and absorption spectra using a Monte-Carlo
sampling technique.  For the other species, a $\pm$20\% error of the column densities was assumed.
The uncertainty is marked by the blue shading around the curves.} } \label{coldens}
\end{figure*}

In principle, the differential rotation of the Milky Way separates spectral features at different Galactocentric radii into distinct locations in velocity
space \citep[see, e.g.,][]{Vallee2008}. Tab.~\ref{tab:distances} lists the different velocity components detectable along the sight-line to the Galactic Center, with the color referring to Fig.~\ref{coldens}. The distance determination and hence the assignment to specific spiral arms is 
complicated due to the streaming motions in the arms \citep{Reid2009}.  Particularly, given the kind of gas the ArH$^+$ line is tracing (see discussion below), we also have to allow for the possibility of detecting inter-arm gas. The exact location of the Galactic center gas observed toward the SgrB2 sources within the \new{Central Molecular Zone} is not easily established, due to the non-circular orbits in the Bar potential \citep{RF2008}. 
\begin{table}
 \caption{Mapping of location in the Galaxy to velocity regions and color coding toward SgrB2 (see Fig.~\ref{coldens}).}\label{tab:distances}
\begin{tabular}{lcl}
 \hline
Component &  v$_{\rm LSR}$ & color code\\
& (km s$^{-1}$)&\\
\hline
 Galactic center& -136 to -55 & green\\
 Norma arm &  -50 to -13 & red\\
 Galactic center &  -9 to 8 & green\\
 \new{Sagittarius arm }&  12 to 22 & blue\\
 \new{Scutum arm }&  25 to 39 & orange\\
 Sagittarius B2 & 47 to 89 & light green\\
 \hline
\end{tabular}
\end{table}


\begin{figure}
\centering
\includegraphics[width=9cm]{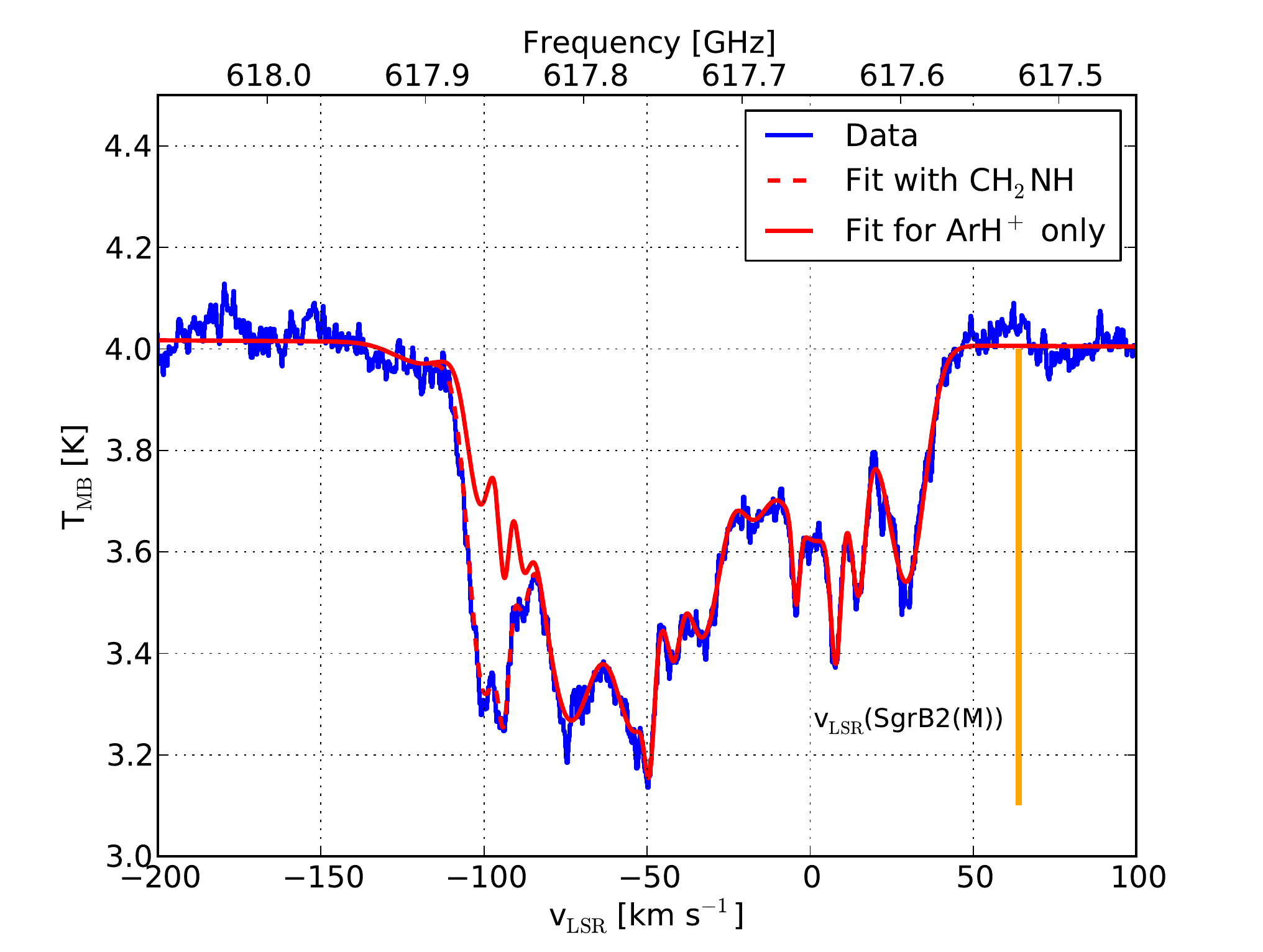}
\caption{Spectrum of  $^{36}$ArH$^+$(1--0) toward SgrB2(M), with fit \vn{including the H$_2$CNH line blending at --110 km s$^{-1}$ (see Fig.~\ref{ch2nh}) as a dashed red line, and fit of $^{36}$ArH$^+$ only in red}.  \st{The mismatch at --110 km/s comes from the H$_2$CNH line (see Fig.~\ref{ch2nh}), which was used in the fit, but what is shown is just the $^{36}$ArH$^+$.}  Note the lack of absorption  at the source velocity 64 km/s, indicated by the vertical orange marker.}  \label{arh+-sgrb2m-fit}
\end{figure}

\begin{figure}
\centering
\includegraphics[width=9cm]{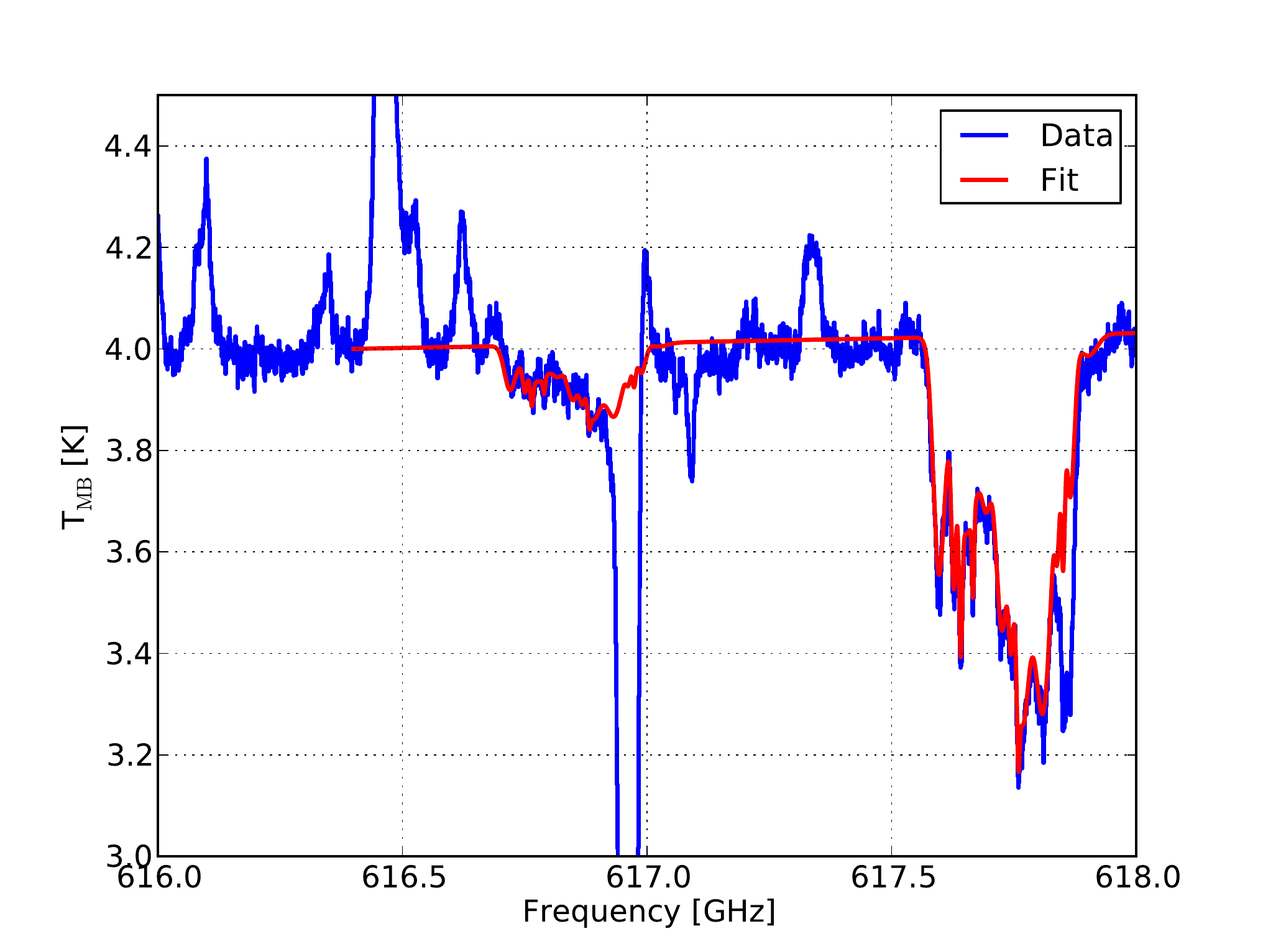}
\includegraphics[width=9cm]{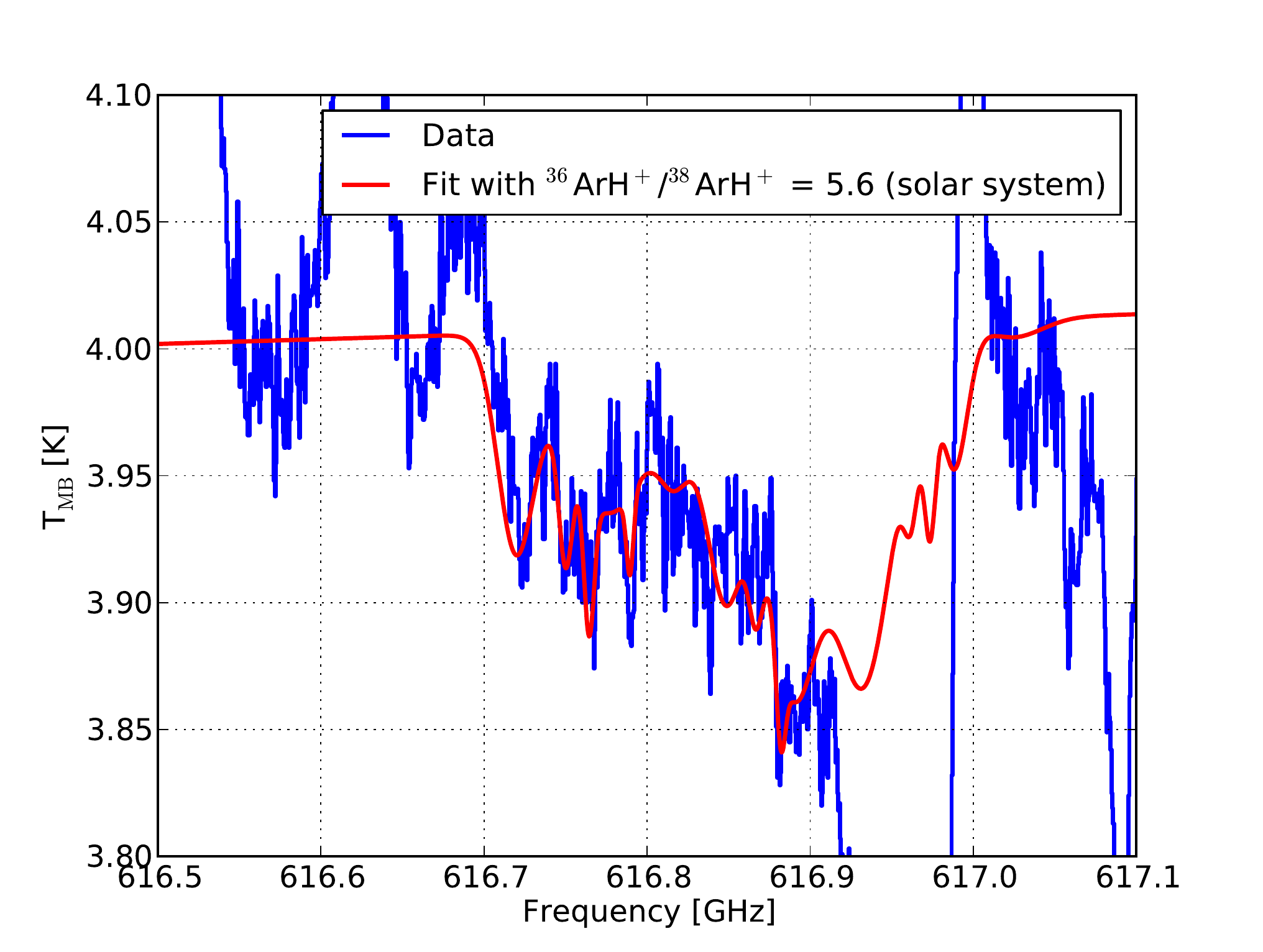}
\caption{Spectra with predictions for $^{36}$ArH$^+$ and $^{38}$ArH$^+$ (top) and zoom in to $^{38}$ArH$^+$ (bottom). The $^{38}$ArH$^+$ spectrum is scaled from  $^{36}$ArH$^+$ assuming as $^{36}$Ar/$^{38}$Ar ratio \st{of 3.3. In the bottom panel, a synthetic spectrum with} the solar system value of 5.6.\st{ is also shown.}} \label{both-arh+-fit}
\end{figure}

\citet{Barlow2013} detected the (1--0) and (2--1) transitions of $^{36}$ArH$^+$ in emission toward the Crab Nebula, the remnant of supernova 1054. The OH$^+$ ion was detected in the same spectra, and both ions are probably excited mainly by warm electrons in the same filaments and knots that show low-ionization atomic lines in the visible spectrum and H$_2$ emission lines in the infrared. The conditions in the general ISM we observe, and hence the chemistry, are very different.  

\new{In SgrB2(M),} we detect only one very wide line, in absorption, which does not absorb at the source intrinsic velocity (see Fig.~\ref{arh+-sgrb2m-fit}). The breadth of the absorption features introduces a potential uncertainty in the identification, which is definitively resolved by the observation of two different isotopic forms of ArH$^+$.  Our non-detection of the (2--1) line, which is also covered by the survey, is not surprising, since the molecule possesses a very high dipole moment, thus the transitions have a high critical density, and 
therefore, in the absence of high density or a very strong FIR field, most of the population will be in the rotational ground state.  Fortunately, the $^{38}$ArH$^+$(1--0) line is covered in the observation as well, 
and, although blended with the CH$_3$OH(4$_{-2,3}      - 3_{-1, 3}$) absorption line, mirrors the absorption pattern of the $^{36}$ArH$^+$(1--0) so closely that 
we do not have any doubt about the correct identification of the 
carrier as $^{36}$ArH$^+$ (Fig.~\ref{both-arh+-fit}).  \vn{The fit was conducted with the solar system value of 5.6, which seems to reproduce the $^{38}$ArH$^+$(1--0) line well.} \st {The derived \new{isotopic} ratio of 3.3 is lower \new{than} the solar system value of 5.6, but the signal-to-noise ratio and the uncertain contamination by the methanol line prevent determining this value with high confidence.} Both $^{36}$Ar and $^{38}$Ar are mostly produced in explosive nucleosynthesis through oxygen burning \citep{Woosley2002}.  \st{We were unable to locate any information on varying $^{36}$Ar/$^{38}$Ar ratios, neither observational nor theoretical.} 

\begin{figure}
\centering
\includegraphics[width=9cm]{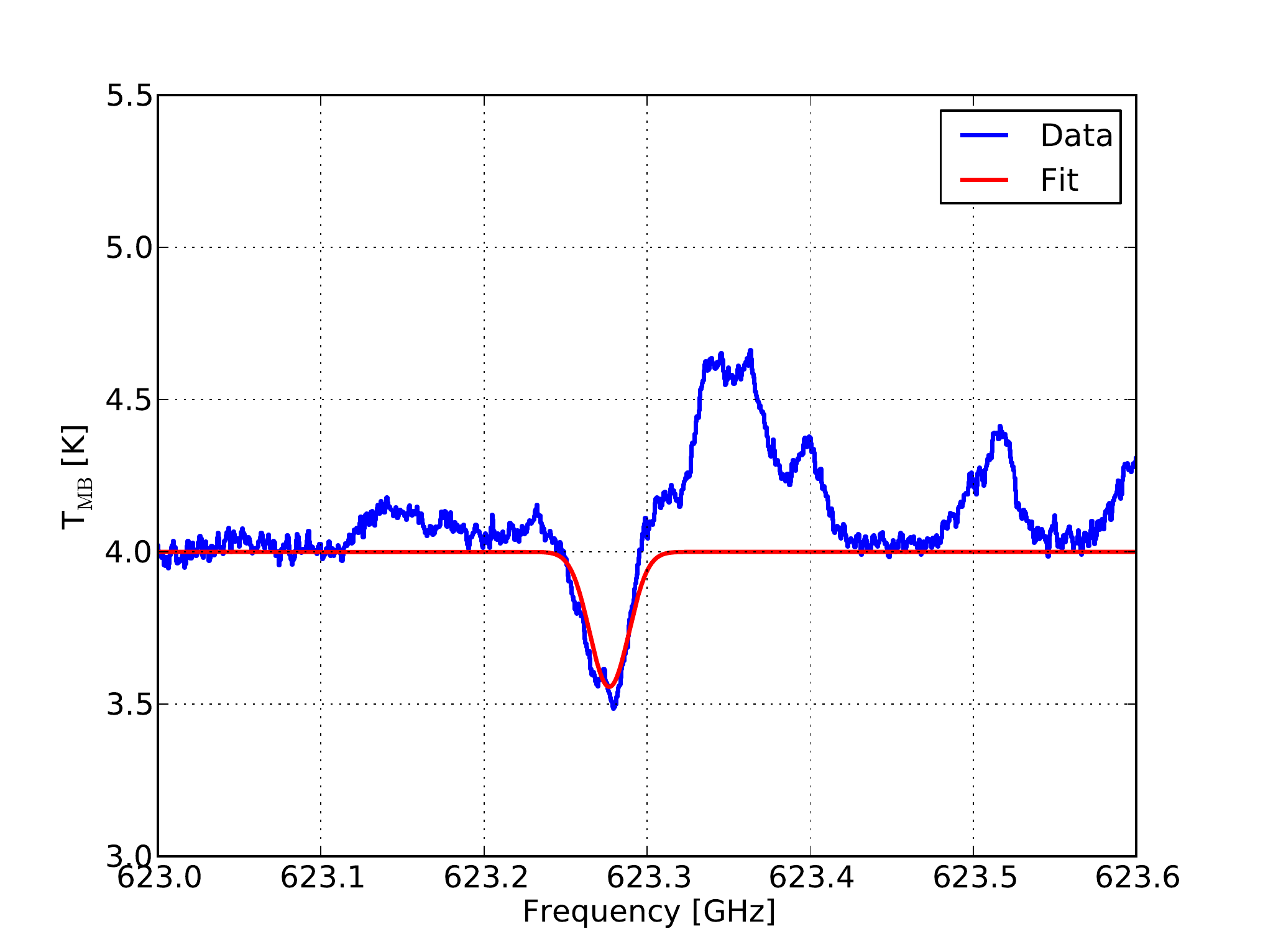}
\includegraphics[width=9cm]{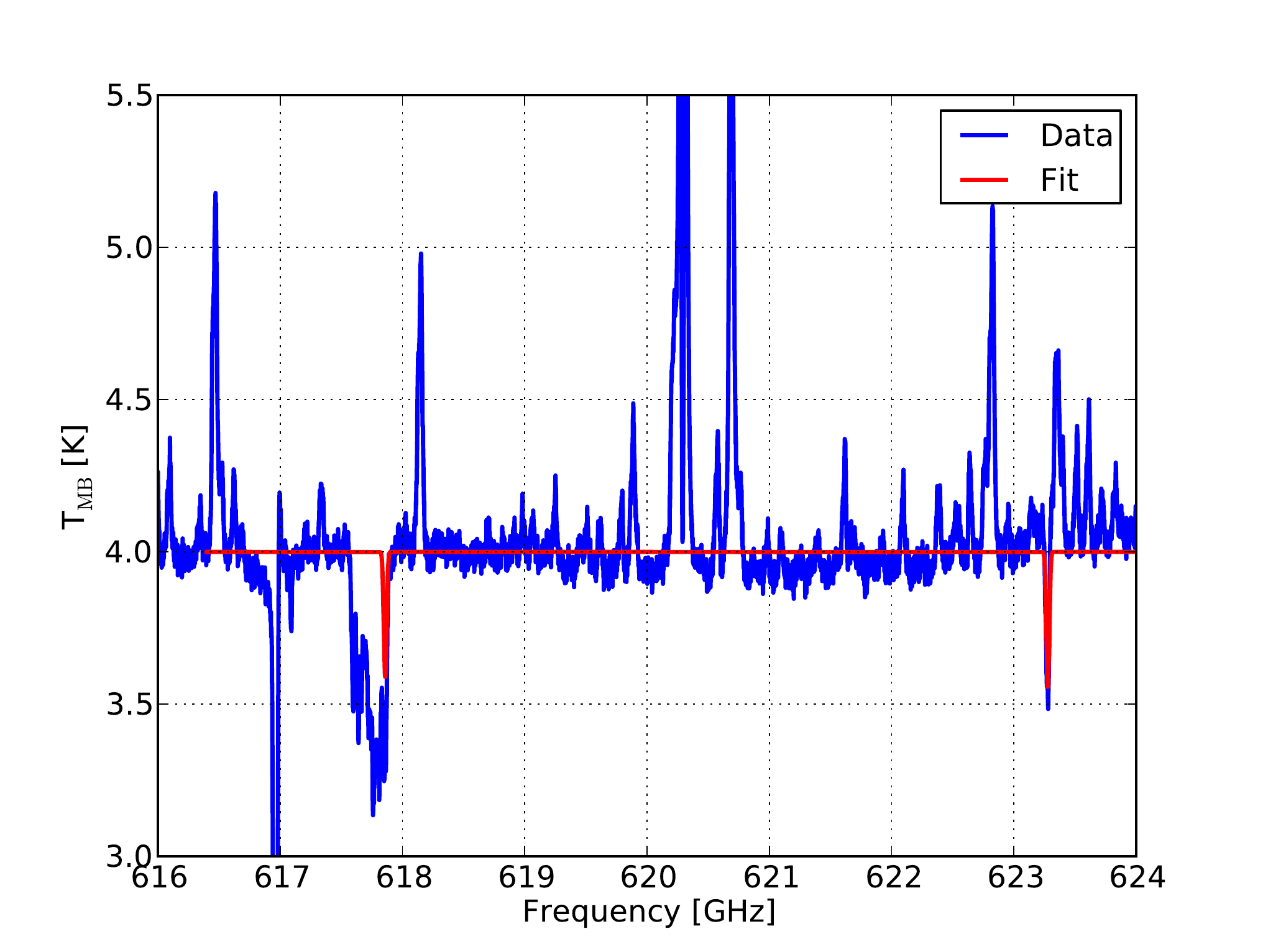}
\caption{The H$_2$CNH($2_{2,0}-1_{1,1}$) line with fit (top) and the two H$_2$CNH lines (bottom). } \label{ch2nh}
\end{figure}

The $^{36}$ArH$^+$ line is blended (at a velocity of about {--110~km s$^{-1}$}) with the H$_2$CNH($2_{2,1}-1_{1,0}$) absorption feature at 617.873~GHz (Fig.~\ref{ch2nh}). The strength of this feature can be estimated, since at 623.292~GHz one finds the H$_2$CNH($2_{2,0}-1_{1,1}$) line, which is almost identical in excitation and line strength (Fig.~\ref{ch2nh}).  However, the  H$_2$CNH($2_{2,0}-1_{1,1}$) line seems to be \new{itself} contaminated with the flank of the adjacent CH$_3$NH$_2(9_{2,6}-8_{1,6})$ emission line and thus the absorption could well be underestimated.  The strength of the {--110~km s$^{-1}$} ArH$^+$ absorption component should therefore be regarded as a lower limit.  The NeH$^+$(1--0) line at 1039.3~GHz is also covered by the survey and is not detected.  Assuming the same excitation conditions, we get a lower limit of [ArH$^+$]/[NeH$^+$] $\ge$ 10.  Although Ne is about 30 times more abundant than Ar, this result is not unexpected, given the different ionization potentials of these 
species (see Sect.
~\ref{chemistry}).

The column densities were determined using the XCLASS{@}CASA program\footnote{http://www.astro.uni-koeln.de/projects/schilke/myXCLASSInterface} (M\"oller et al, in prep), which fits the absorption spectrum with \new{multiple Gaussian components} in opacity (hence taking the line shape changes due to opacity into account), assuming an excitation temperature of 2.7~K, using MAGIX \citep{Moeller2013}.  

In Fig.~\ref{coldens}, we show the ArH$^+$ column density determined in this way, together with the H$_2$ column density, determined from HF absorption, the H$_2$O$^+$ column density, which traces diffuse gas with an H$_2$/H fraction of 5-10\% \citep{Neufeld2010}, atomic hydrogen from Winkel et al.~(in prep.), and the ArH$^+$ abundance relative to atomic hydrogen.  The latter is justified by our result from Sect.~\ref{chemistry}, which shows that ArH$^+$ traces gas with a H$_2$/H of $\approx 10^{-3}$.  The plots are ordered in descending abundance of the species with respect to  H$_2$.  ArH$^+$ does not correlate  either with molecular gas traced by HF, or the diffuse gas traced by H$_2$O$^+$, which points to different molecular fractions of gas traced by ArH$^+$ (f(H$_2) \approx 10^{-3}$) and H$_2$O$^+$ (f(H$_2) \approx 0.1$).  

The ArH$^+$ abundance varies between \st{$3\times 10^{-9}$} \vn{$3\times 10^{-8}$} and  $5\times 10^{-11}$, except at very strong H{\sc i} peaks, and seems to vary smoothly with velocity, with the highest values achieved at the lowest velocities -- which however can be affected by the blending with H$_2$CNH.  Both ArH$^+$ and H$_2$O$^+$ show a distribution in velocity \new{that is} more or less continuous and -- unlike HF -- does not show any breaks \new{associated} with the different spiral arm/Galactic Center velocity components.  This \new{velocity structure} points to the gas responsible for the  ArH$^+$ and H$_2$O$^+$ not only tracing spiral arms, but a more continuous mass distribution including interarm gas, which appears to be less molecular \citep{Sawada2012a,Sawada2012b}.  

\begin{figure}
\centering
\includegraphics[width=9cm]{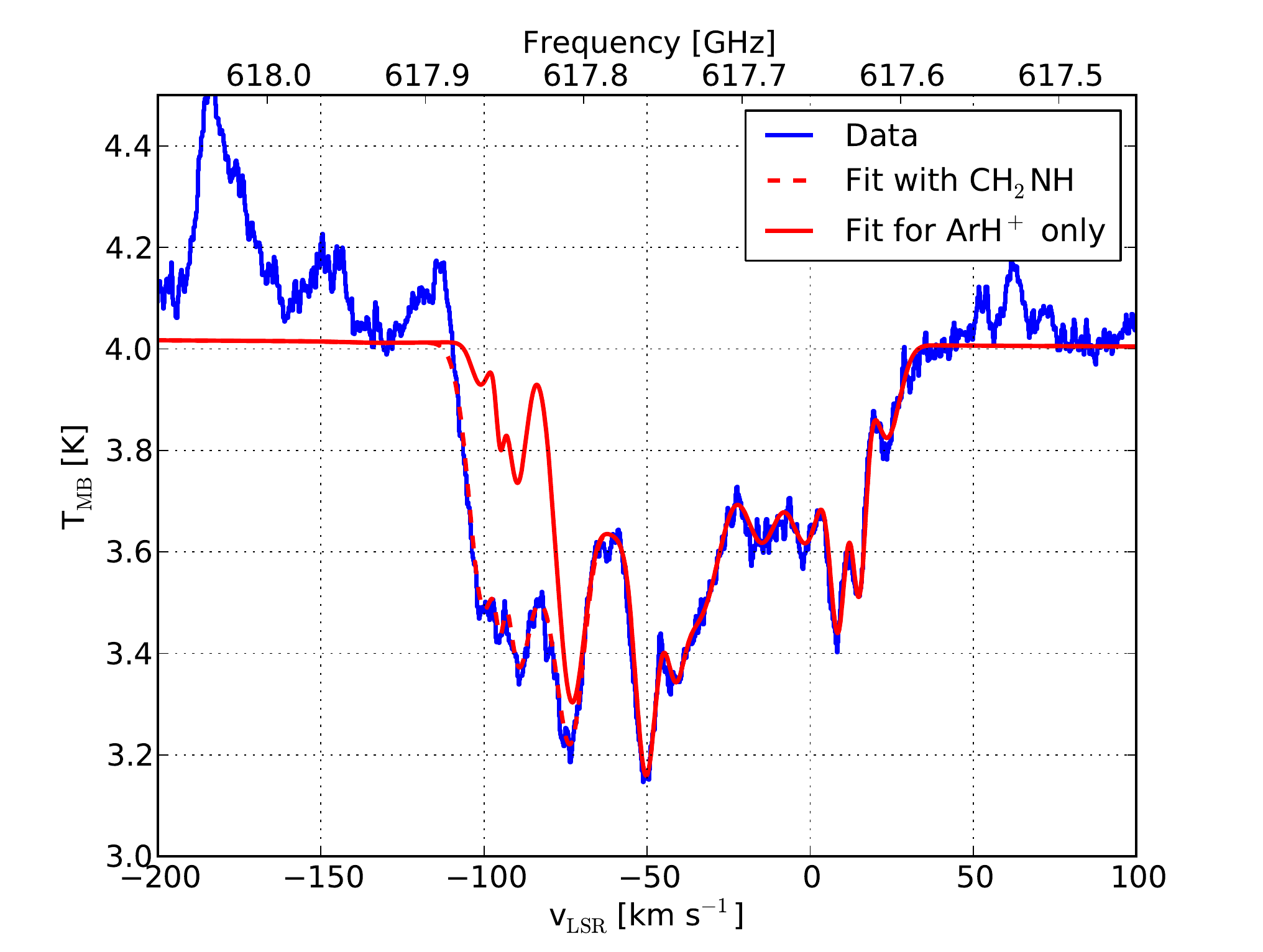}
\caption{Fit of SgrB2(N).  The H$_2$CNH($2_{2,0}-1_{1,1}$) lines (here with the additional 80 km s$^{-1}$ component) has been taken out the same way as for SgrB2(M).  There has been no correction for emission lines from the background source, which most likely distort the absorption profile.} \label{arh+-sgrb2n-fit}
\end{figure}

In Fig.~\ref{arh+-sgrb2n-fit}, we show the fit toward SgrB2(N).  It \vn{morphologically} looks very different from that toward SgrB2(M), which opens the exciting possibility to study the variation of the ArH$^+$ column density between two nearby lines-of-sight, as have been seen for H$_3$O$^+$ toward the same sightlines \citep{Lis2014}.  However, there is still contamination by emission lines from the background source, which is more pronounced for SgrB2(N) than for SgrB2(M).  An investigation of all emission lines from the survey, which is planned, would enable us to predict the background source emission, and therefore the variation of ArH$^+$.  At present, however, this investigation has not yet been completed, and in the absence of solid evidence we take the prudent approach of assuming that most of the variations are due to emission line contamination.

\begin{figure*}
\centering
\includegraphics[width=18cm]{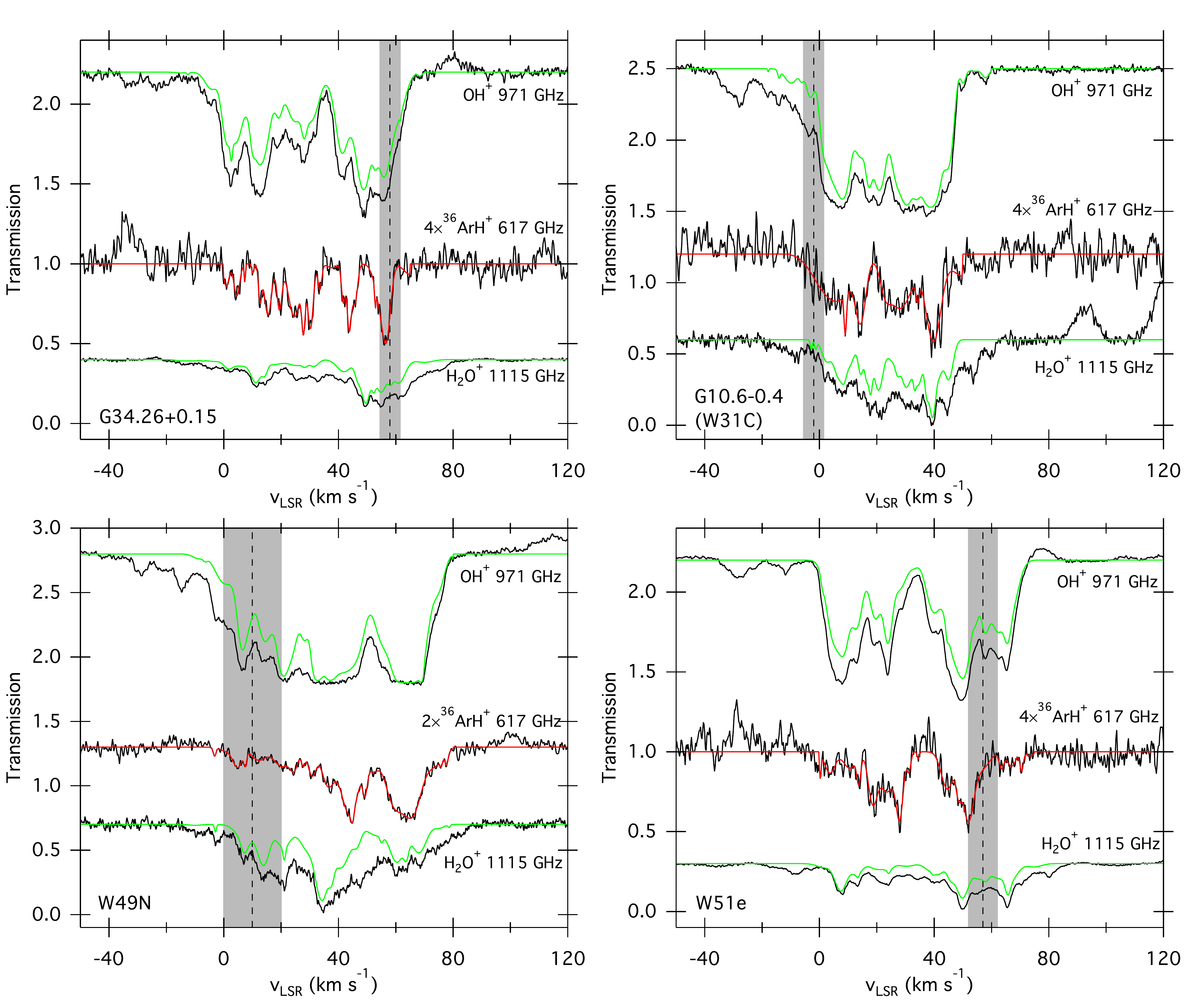}
\caption{Observations of OH$^+$, H$_2$O$^+$, and ArH$^+$ toward PRISMAS sources. The green lines for OH$^+$ and H$_2$O$^+$, whose transitions have hyperfine structure,  give the deconvolved strength of the main hyperfine component.The red lines are fits to the ArH$^+$ spectra used in calculating the column densities presented in Figure 7. Note that ArH$^+$ spectra have been scaled up to more clearly show the absorption profiles. Vertical dashed lines and gray shaded regions mark the systemic velocity and velocity dispersion observed for background sources.  All OH$^+$ and H$_2$O$^+$ spectra in these sight lines will be presented and analyzed in detail by Indriolo et al. 2014 (in prep.).  Analyses of OH$^+$ and H$_2$O$^+$ that only utilized a subset of the eventual data have been performed for W31C \citep{OH_n^+_2010}, W49N \citep{Neufeld2010}, and W51 \citep{OH+_det_2010, CRI-rate_W51_2012}.} \label{prismas-data}
\end{figure*}

\begin{figure*}
\centering
\includegraphics[width=9cm]{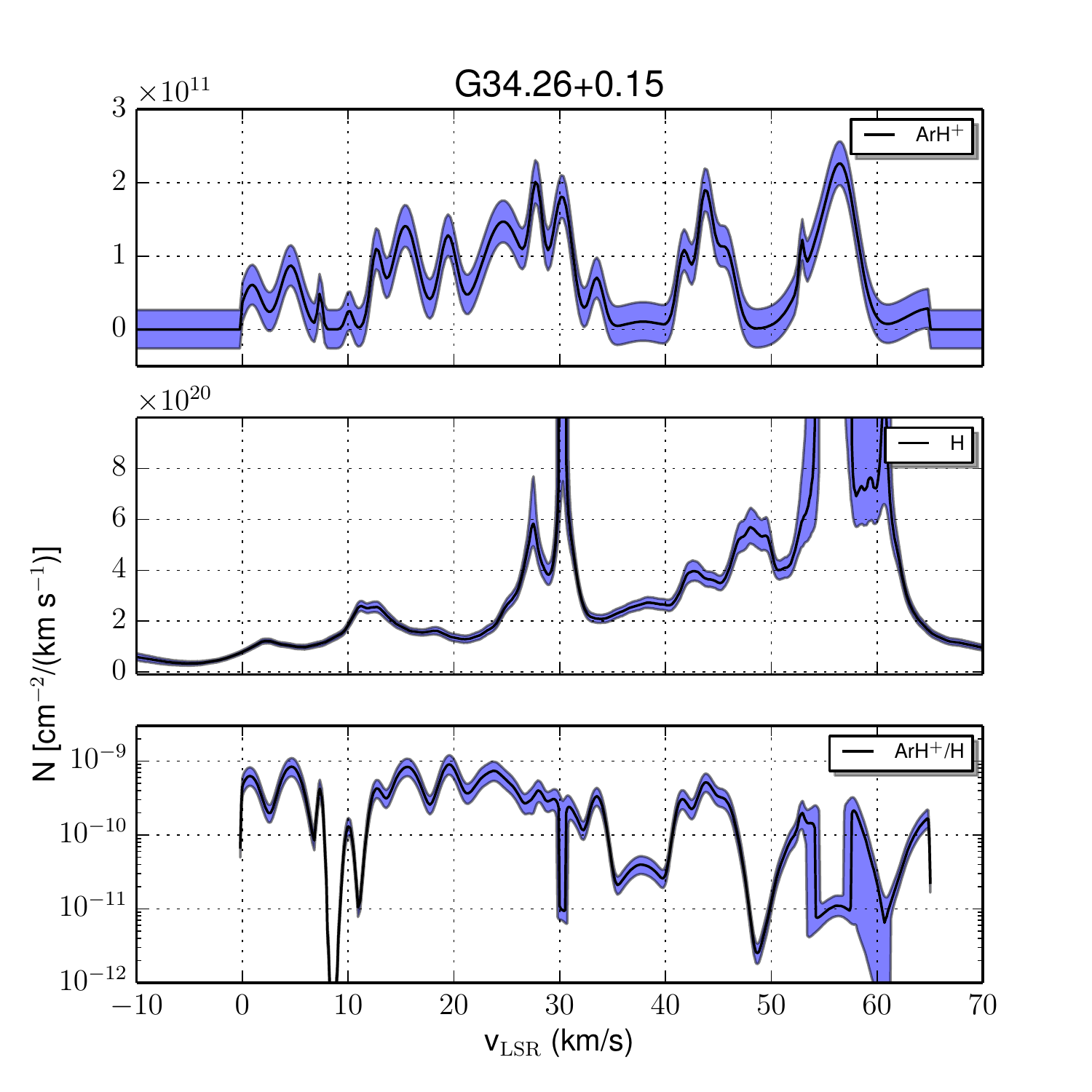}
\includegraphics[width=9cm]{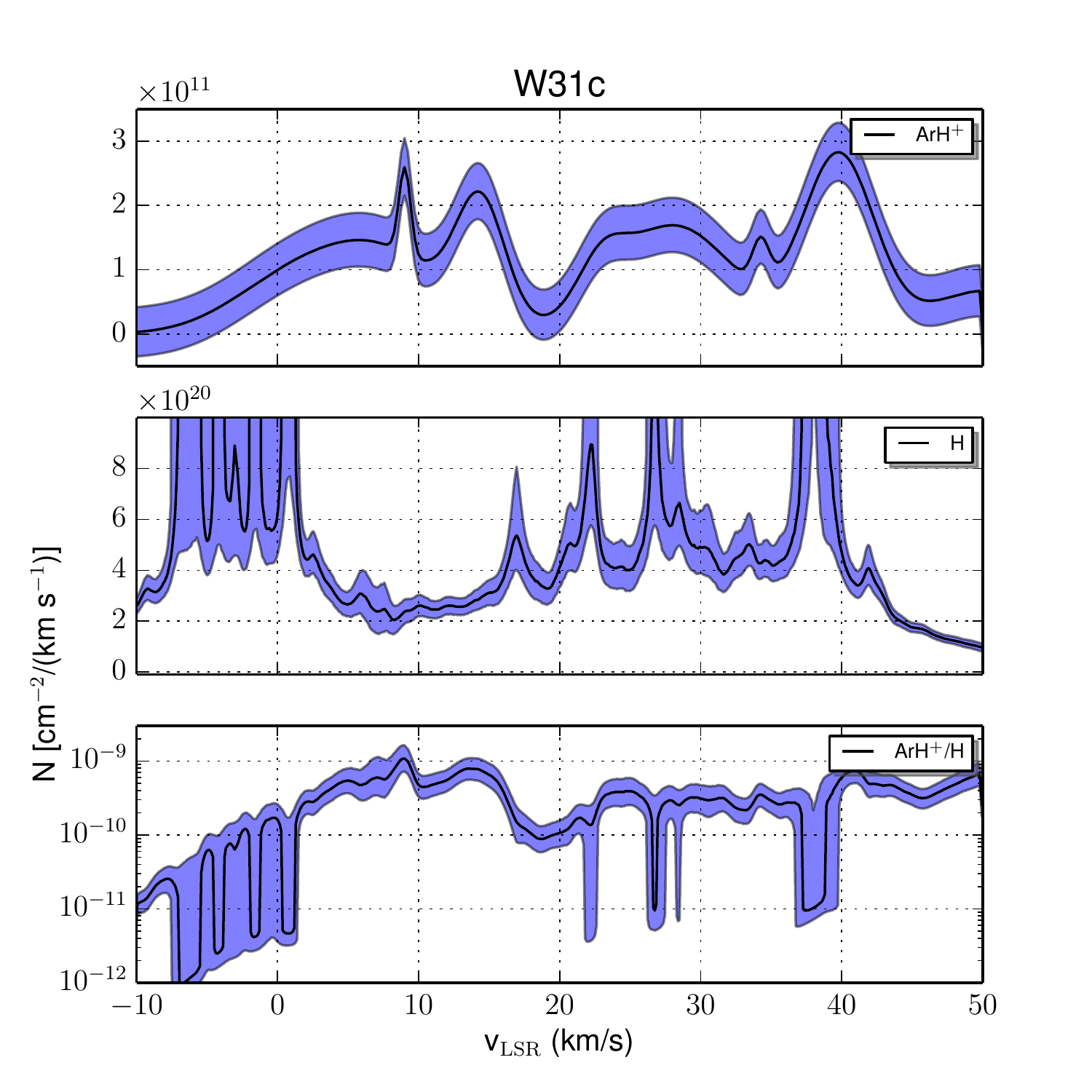}
\includegraphics[width=9cm]{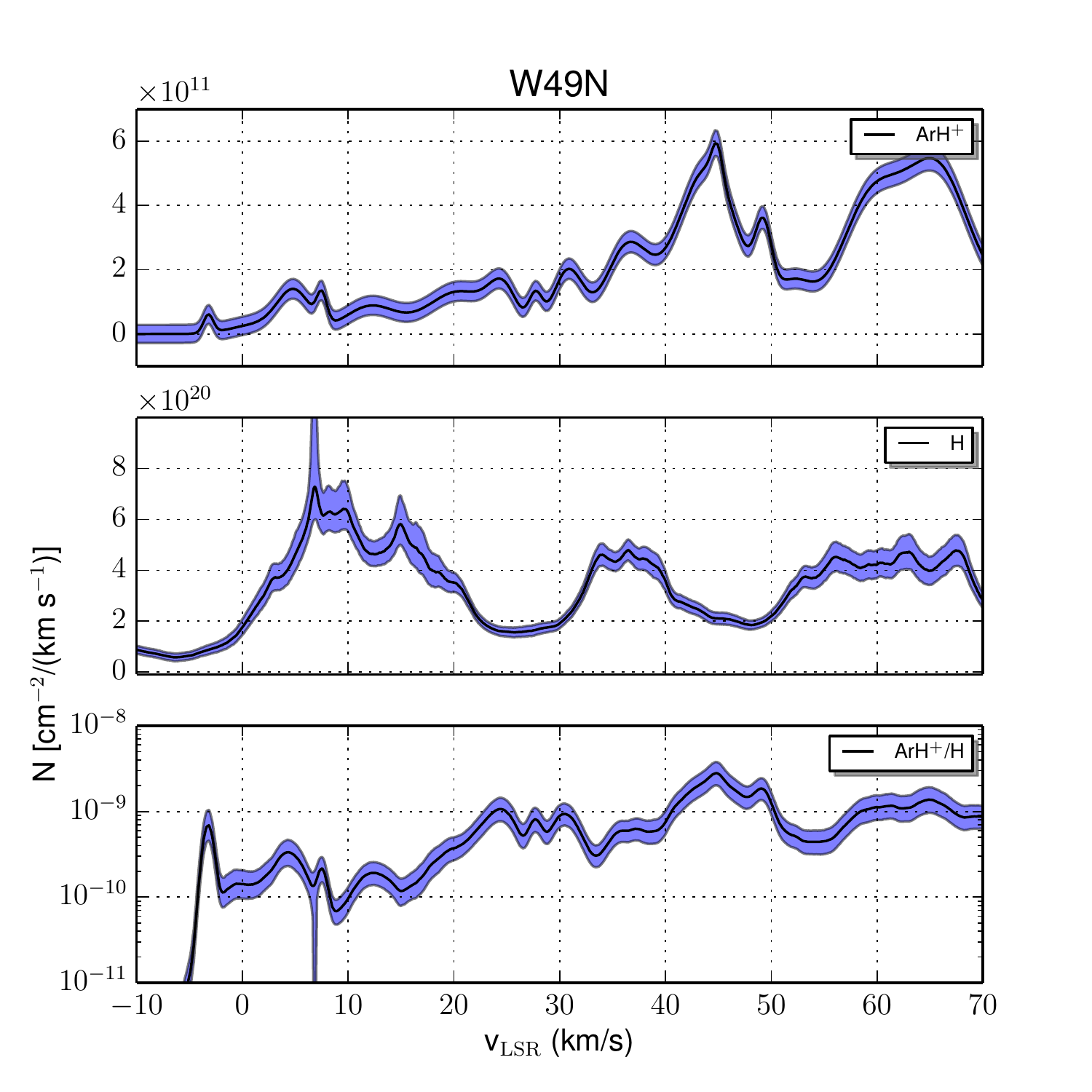}
\includegraphics[width=9cm]{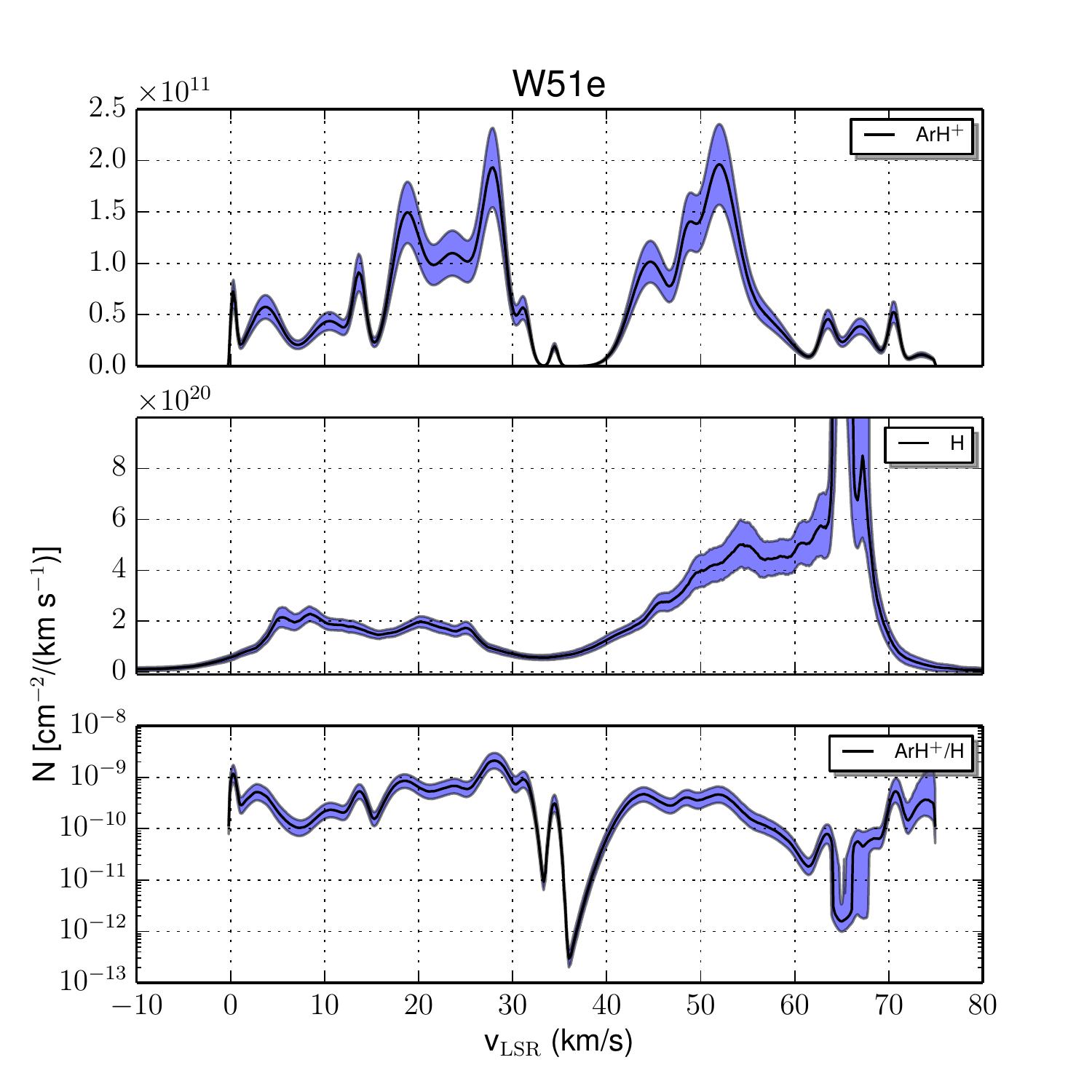}
\caption{Column density per km s$^{-1}$ of ArH$^+$ and H and abundance of ArH$^+$ relative to H toward the PRISMAS sources.  The H column densities come from Winkel et al. (in prep). \new{The uncertainty is marked by the blue shading around the curves.}} \label{prismas-fits}
\end{figure*}

In Fig.~\ref{prismas-data} and \ref{prismas-fits} we show the data obtained toward the PRISMAS sources G34.25+0.15, W31C (G10.6), W49N, and W51e, together with ArH$^+$  and H column densities (Winkel et al., in prep) and ArH$^+$ abundances relative to H.  ArH$^+$ is \new{weak or absent} toward the source envelopes. In the following, we briefly describe the individual sources, following the discussion in \citet{Godard2012} and \citet{Flagey2013}. 
\paragraph{G34.26+0.15} has a source intrinsic velocity of v$_{\rm LSR}$ = 58 km~s$^{-1}$. The ArH$^+$ absorption at $\approx$60 km~s$^{-1}$ is associated with a strong absorption
feature tracing infalling material.
Foreground gas is detected at velocities between 0 and 45 km~s$^{-1}$.
\paragraph{W31C} has a source intrinsic velocity of v$_{\rm LSR}$ = --2 km~s$^{-1}$. Foreground gas is detected between $\approx $10 and 50 km~s$^{-1}$. \new{The strongest feature appears at ~ 40km~s$^{-1}$ , at the
same velocity as H$_3$O$^+$ absorption.}
There could be a weak and broad ArH$^+$ absorption associated with this source.
\paragraph{W49N} has a source intrinsic velocity of v$_{\rm LSR}$ = 10 km~s$^{-1}$.  \new{This sight-line presents the strongest ArH$^+$ absorption outside
the Galactic center. Given the large distance (11.4 kpc). the line of sight
crosses two spiral arms. The absorption is stronger in the 65 km~s$^{-1}$
feature, associated with the Sagittarius spiral arm}.  The weak absorption near 10 km~s$^{-1}$ may be associated with W49 itself.
\paragraph{W51e} has a source intrinsic velocity of v$_{\rm LSR}$ = 57 km~s$^{-1}$. Foreground absorption appears between 0 and 45 km~s$^{-1}$
and there is a deep absorption near 65 km~s$^{-1}$ associated with an infalling layer
in the W51 complex.
The gas near 22 km~s$^{-1}$ is prominent in CH$^+$ and C$^+$, not in molecular lines
but shows up weakly in HF and H$_2$O. Some of the H{\sc i} signal could be associated with the outflow in W51e.

The spectra also show that there is no very tight correlation with the gas traced by OH$^+$ and H$_2$O$^+$.  The ArH$^+$ abundances relative to H are similar to those measured on the  SgrB2 sightlines, viz.\ $3\times 10^{-9}$ and  $10^{-11}$. The H{\sc i} data toward all sources have some high column density spikes that are probably
artifacts related to high opacity regions and the corresponding ArH$^+$ abundance should be disregarded. \new{The continuity of the ArH$^+$ absorption and the large width may indicate that some features are associated with the interarm gas.}

\section{Chemistry of argon in diffuse interstellar clouds}\label{chemistry}
\subsection{Basic features of argon chemistry}

The interstellar chemistry of the element argon shows several noteworthy features that we list below.  

\medskip

1) The ionization potential of atomic argon, {\it IP}(Ar) = 15.76~eV, is greater than that of hydrogen, {\it IP}(H) = 13.5986~eV.  As a result, argon is shielded from ultraviolet radiation capable of ionizing it, and is primarily neutral in the cold neutral medium.

\medskip

2) The proton affinity of argon, {\it PA}(Ar) = 369~kJ~mol$^{-1}$ (Hunter \& Lias 1998), is smaller than that of molecular hydrogen, {\it PA}(H$_2$)=422~kJ~mol$^{-1}$.  As a result, proton transfer from H$_3^+$ to Ar is endothermic, with an endothermicity equivalent to 6400~K (Villinger et al.\ 1982).  Moreover, the argonium ion, ArH$^+$, can be destroyed by means of an exothermic proton transfer to H$_2$, as well to other neutral species with proton affinities greater than that of Ar: these include C, N, CO, and O (Rebrion et al.\ 1989; Bedford \& Smith 1990).  Most importantly, however, the proton affinity of Ar is larger than that of atomic hydrogen: thus, ArH$^+$ is not destroyed by reaction with atomic hydrogen in the cold diffuse medium.

\medskip

3) The ionization potential of atomic argon, {\it IP}(Ar) = 15.76~eV, is {\it smaller} than {\it IP}(H) + $D_0({\rm H}_2$) = 18.09~eV, where $D_0({\rm H}_2)$ = 4.48~eV is the dissociation energy of H$_2$.  As a result, the dissociative charge transfer reaction ${\rm Ar}^+ + {\rm H}_2 \rightarrow {\rm Ar} + {\rm H} + {\rm H}^+$ is endothermic and negligibly slow at the temperature of diffuse interstellar clouds.  Thus, in the reaction of Ar$^+$ and H$_2$, the primary product channel leads to the formation of 
ArH$^+$ via the H atom abstraction reaction ${\rm Ar}^+ + {\rm H}_2 \rightarrow {\rm ArH}^+ + {\rm H}$.  This thermochemistry is different from that of the more abundant noble gas elements He and Ne, which have ionization potentials (24.5874~eV and 21.5645~eV respectively) in excess of {\it IP}(H) + $D_0({\rm H}_2)$; those elements do not efficiently form a hydride cation through reaction of their cation with H$_2$, because the product channel is dominated by dissociative charge transfer.

\medskip

4) Dissociative recombination (DR) of ArH$^+$ (i.e., ${\rm ArH}^+ + e \rightarrow {\rm Ar} + {\rm H}$) is unusually slow. While almost all diatomic molecular ions, including HeH$^+$ and NeH$^+$ (Takagi 2004; Mitchell et al.\ 2005a), undergo rapid dissociative recombination (DR) at the temperatures of diffuse clouds  ($\simlt 100$~K) -- with typical rate coefficients $\sim 10^{-7}{\rm \, cm}^{3}\,{\rm s}^{-1}$ -- recent storage ring measurements of the DR of ArH$^+$ have found the process too slow to measure at energies below 2.5~eV (Mitchell et al.\ 2005b).  While peaks in the DR cross-section have been found at electron energies of 7.5, 16, and 26 eV, and are readily understood with reference to the potential energy curves for ArH$^+$, these higher energies are not relevant in cold interstellar gas clouds.  Thus, the experimental data place an upper limit of $10^{-9}{\rm \, cm}^{3}\,{\rm s}^{-1}$ on the DR rate coefficient at interstellar temperatures.

\medskip

5) The photodissociation rate for ArH$^+$ is unusually small.  At wavelengths beyond the Lyman limit (i.e., $> 912 \AA$) photodissociation is dominated by transitions to a repulsive B$^1\Pi$ state, with a vertical excitation energy of 11.2~eV, and to  a repulsive A$^1\Sigma^+$ state, with an excitation energy of 15.8~eV (Alexseyev et al.\ 2007).  The former transition has an unusually small dipole moment (0.13~D), while the latter provides a strong absorption feature that has its peak shortward of the Lyman limit.  Recent theoretical calculations of the photodissociation cross-section have been performed by Alexseyev et al.\ (2007) using the multireference Spin-Orbit Configuration Interaction approach.  Adopting this cross-section, we estimate a photodissociation rate of only $1.0 \times 10^{-11}\,{\rm \, s}^{-1}$ for an unshielded ArH$^+$ molecule exposed to
the mean interstellar radiation field (ISRF) given by Draine (1978). This value is more than two orders of magnitude smaller than that for the isoelectronic HCl molecule (Neufeld \& Wolfire 2009).  \new{ A similar estimate of the ArH$^+$ photodissociation rate was obtained from the same theoretical cross-sections by Roueff et al. (2014, in press) }

\medskip

6) The primary cosmic ray ionization rate for Ar is an order of magnitude larger than that for H (Kingston 1965; Jenkins 2013).   

\medskip

As is always the case in interstellar chemistry (Neufeld \& Wolfire 2009; hereafter NW09), basic thermochemical facts (i.e., 1 through 3 above) play a key role.   Clearly, (1) and (2) above are detrimental to the production and survival of argonium in the interstellar medium, while (3) enhances the production rate relative to HeH$^+$ and NeH$^+$.  Two unusual features of the kinetics of ArH$^+$ (4 and 5) enhance its survival in the diffuse ISM, while consideration (6) enhances the production of Ar$^+$ relative to that of H$^+$.

\subsection{Diffuse cloud models}\label{sec:chem}

In modeling the chemistry of argonium in diffuse molecular clouds, we have modified the diffuse cloud model presented by NW09 and Hollenbach et al.\ (2012) by the addition of the reactions listed in Table~\ref{tab:chem}.  In this reaction network, ArH$^+$ is produced in a two step process, in which  atomic argon undergoes ionization by cosmic rays, and the resultant Ar$^+$ ion reacts with H$_2$ to form ArH$^+$.  ArH$^+$ is destroyed by photodissociation, or by transferring a proton to a neutral species (primarily O or H$_2$) of larger proton affinity than Ar. \new{We note here that in our model we cannot distinguish if the primary ionization is caused by cosmic rays or X-rays, which can play a role in the Galactic center.}

\begin{table}

\caption{Reaction list}\label{tab:chem}
\begin{tabular}{llr}

\hline
\\
Reaction & Assumed rate or rate coefficient & Notes\\
\\
\hline
\\
$\rm Ar +CR \rightarrow Ar^+ + e$ & $(10 + 3.85 \phi) \zeta_p({\rm H}) $ & (1) \\
$\rm Ar + H_2^+ \rightarrow Ar^+ + H_2$ & $10^{-9}\,\rm cm^{3}\,s^{-1} $ & (2)  \\
$\rm Ar + H_3^+ \rightarrow ArH^+ + H_2$ & $8 \times 10^{-10}\exp(-6400\,{\rm K}/T)\,\rm cm^{3}\,s^{-1} $ & (3)  \\
$\rm Ar^+ + e \rightarrow Ar + h\nu$ & $3.7 \times 10^{-12}\,(T/300\,{\rm K})^{-0.651} \rm cm^{3}\,s^{-1} $ & (4)  \\
$\rm Ar^+ + PAH^- \rightarrow Ar + PAH$	& $6.8 \times 10^{-8}  \,(T/300\,{\rm K})^{-0.5}\,\rm cm^{3}\,s^{-1} $ & (5)  \\
$\rm Ar^+ + PAH \rightarrow Ar + PAH^+$	& $5.9 \times 10^{-9}  \,\rm cm^{3}\,s^{-1} $ & (5)  \\
$\rm Ar^+ + H_2 \rightarrow ArH^+ + H$	& $8.4 \times 10^{-10} \,(T/300\,{\rm K})^{0.16}\,\rm cm^{3}\,s^{-1} $ & (6)  \\
$\rm ArH^+ + H_2 \rightarrow Ar + H_3^+$	& $8 \times 10^{-10} \,\rm cm^{3}\,s^{-1} $ & (3)  \\
$\rm ArH^+ + CO \rightarrow Ar + HCO^+$	& $1.25 \times 10^{-9} \,\rm cm^{3}\,s^{-1} $ & (3)  \\
$\rm ArH^+ + O \rightarrow Ar + OH^+$	& $8 \times 10^{-10} \,\rm cm^{3}\,s^{-1} $ & (7)  \\
$\rm ArH^+ + C \rightarrow Ar + CH^+$	& $8 \times 10^{-10} \,\rm cm^{3}\,s^{-1} $ & (7)  \\
$\rm ArH^+ + e \rightarrow Ar + H$	    & $< 10^{-9} \,\rm cm^{3}\,s^{-1} $ & (8)  \\
$\rm ArH^+ + h\nu \rightarrow Ar^+ + H$	& $ 1.0 \times 10^{-11} \chi_{\rm UV}\,f_A \,s^{-1} $ & (9)  \\
\\
\hline
\\
\end{tabular}

\scriptsize
\noindent (1) Kingdon (1965); Jenkins (2013); $\phi$ is the number of secondary ionizations of H per primary cosmic ray ionization: we adopt the fit given by Dalgarno et al.\ (1999)

\noindent (2) Estimate

\noindent (3)	Villinger et al.\ (1982) 

\noindent (4)	Shull \& van Steenberg (1982)

\noindent (5)	Hollenbach et al.\ (2012), scaled by reduced mass$^{-0.5}$

\noindent (6)	Rebrion et al.\ (1989); Bedford and Smith (1990) 

\noindent (7)	Assumed equal to the rate for reaction with H$_2$

\noindent (8)	Mitchell et al.\ (2005b)

\noindent (9)	Unshielded rate based on theoretical cross-sections of Alexseyev et al.\ (2007).  Attenuation factor $f_A = [E_2(3.6 A_V)+ E_2(3.6 [A_V({\rm tot}) - A_V])]/2$, where $E_2$ is an exponential integral.  (Based on attenuation factor obtained by NW09 for photoionization of Cl.)

\end{table}

\begin{figure}
\centering
\includegraphics[width=9cm]{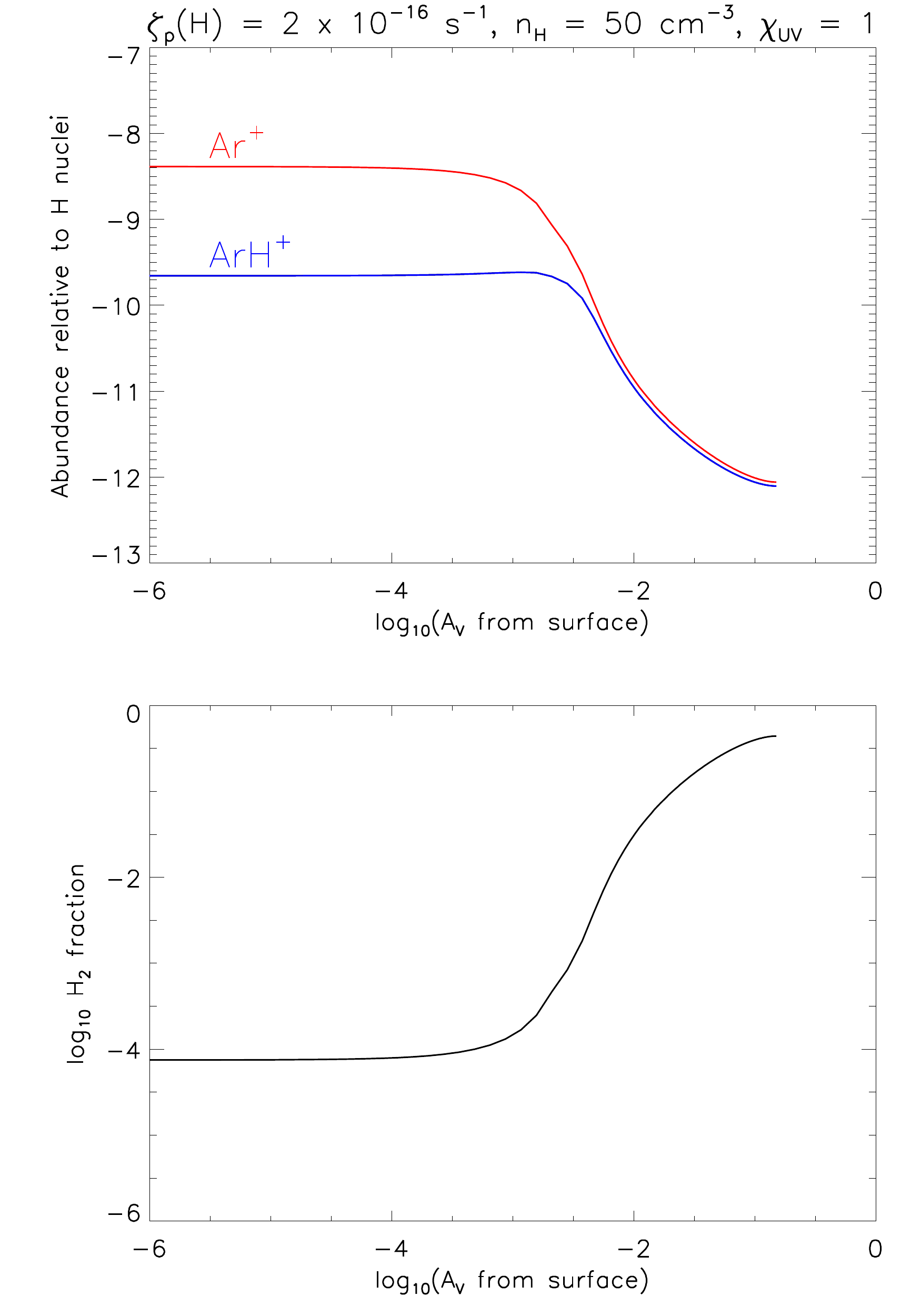}
\caption{Abundances of Ar$^+$ and ArH$^+$ as function of A$_V$ (upper panel) and f(H$_2$) (lower panel). The elemental abundance of Ar is $3.2\times 10^{-6}$, so most of the Argon is still neutral.} \label{fig_a1}
\end{figure}

In Figure~\ref{fig_a1} we show the resultant Ar$^+$ and ArH$^+$ abundances for our standard diffuse cloud model.  Here, a cloud with an assumed density, $n_{\rm H}$, of 50 H nuclei per cm$^{-3}$, is modeled as a slab that is irradiated from both sides by a UV radiation field of intensity equal to that of the mean ISRF (Draine 1978).   The assumed primary cosmic ray ionization rate for atomic hydrogen is $\zeta_p({\rm H}) = 2 \times 10^{-16}\,{\rm s}^{-1}$.  Results are shown as a function of depth below the cloud surface, measured in terms of visual extinction, $A_V = 5.9 \times 10^{-22} N_{\rm H}\,{\rm cm}^2$\,mag when $N_{\rm H}$ is the column density in cm$^{-2}$.  For the model shown in Figure~\ref{fig_a1}, the total visual extinction through the slab is $A_V({\rm tot})$=0.3~mag, and thus the slab midplane is located at $A_V$ = 0.15~mag.

\begin{figure}
\centering
\includegraphics[width=9cm]{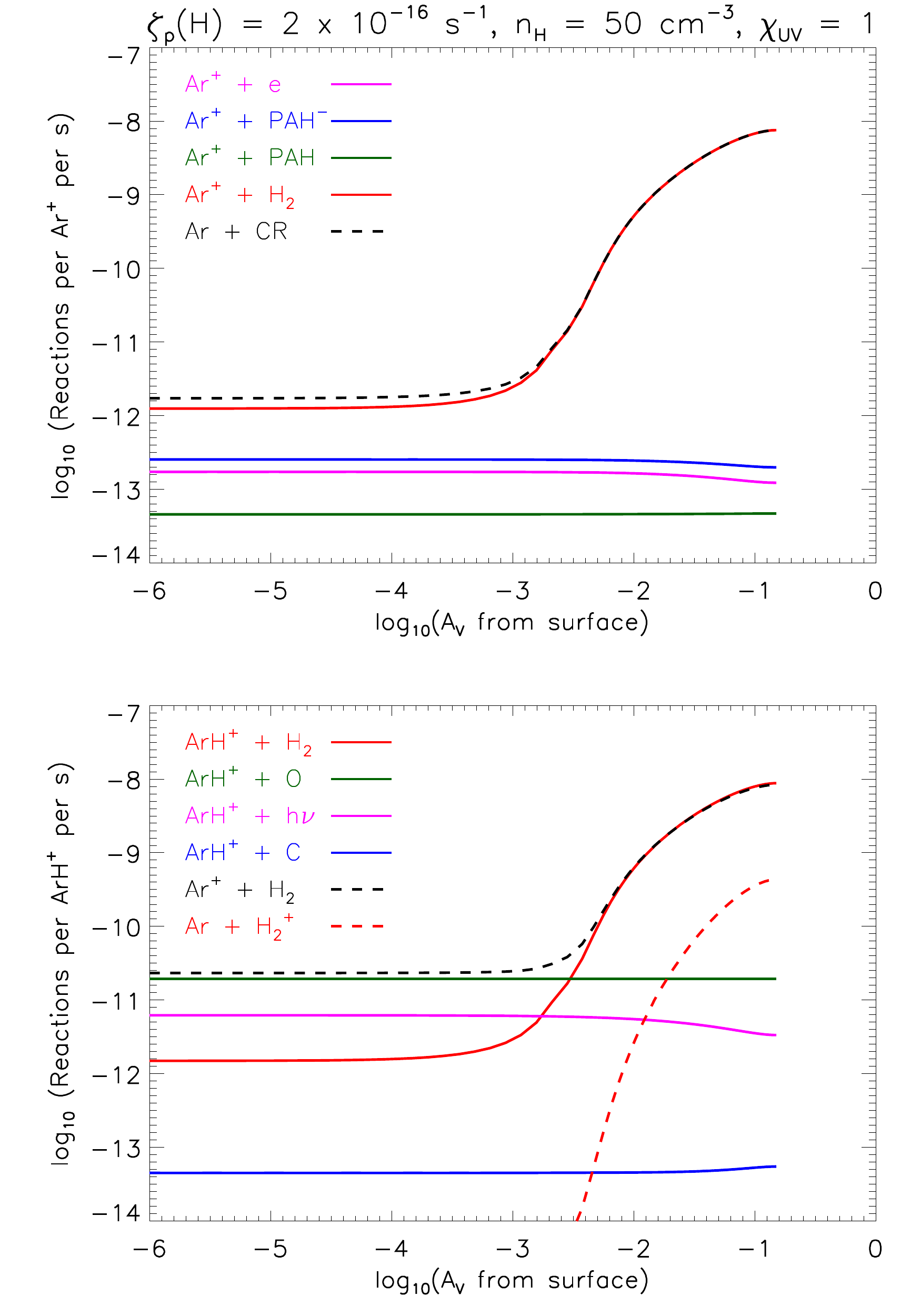}
\caption{Rates of formation (dashed lines) and destruction (solid lines) by various processes are shown for Ar$^+$ (upper panel) and ArH$^+$ (lower panel).} \label{fig_a2}
\end{figure}

In Figure~\ref{fig_a2}, the rates of formation (dashed lines) and destruction (solid lines) by various processes are shown for Ar$^+$ (upper panel) and ArH$^+$ (lower panel).  The upper panel of Figure~\ref{fig_a2} shows that -- even close to the cloud surface where the molecular fraction is smallest -- the destruction of Ar$^+$ is dominated by reaction with H$_2$ to form ArH$^+$.  Competing pathways, including mutual neutralization with PAH cations, charge transfer with neutral PAHs, and radiative recombination are almost negligible for molecular hydrogen fractions $\simgt 10^{-4}$.  Thus, once 
the molecular fraction exceeds $10^{-4}$, more than 75\% of Ar ionizations lead to the formation of ArH$^+$.  

The lower panel of Figure~\ref{fig_a2} indicates that the destruction of ArH$^+$ is dominated by three processes: proton transfer to H$_2$, proton transfer to atomic oxygen, and photodissociation.
Setting the Ar ionization rate equal to the rate of ArH$^+$ destruction via these three processes, we may approximate the predicted ArH$^+$ abundance by the equation
\begin{equation}
{n({\rm ArH}^+) \over n({\rm Ar})} = {\zeta({\rm Ar}) \over k_1 n({\rm O}) + k_2 n({\rm H}_2) + \zeta_{\rm pd} ({\rm ArH}^+)}\label{eq}
\end{equation}
where $k_1 = 8 \times 10^{-10}\,\rm cm^3\,s^{-1}$ is the rate coefficient for proton transfer to H$_2$, $k_2$ (assumed equal to $k_1$) is the rate coefficient for proton transfer to O, and $\zeta_{\rm pd} ({\rm ArH}^+)$  is the photodissociation rate for ArH$^+$ (equal to $1.1 \times 10^{-11}\,\chi_{\rm UV} \rm s^{-1}$ in the limit of no shielding, where $\chi_{\rm UV}$ is the intensity of the ISRF in units of the mean Galactic value given by Draine 1978).  The numerator in equation~(\ref{eq}) is the total ionization rate for Ar, which, following Jenkins (2013), we take as $(10  + 3.85 \phi) \zeta_p(H)$, where $\phi$ is the number of secondary ionizations of H per primary ionization.  Using the fit to $\phi$ given by Dalgarno, Yan \& Liu (1999), we find that $\phi$ ranges from 0.48 to 0.26 within the standard cloud model presented here.  Adopting the middle of that range, we find that $\zeta({\rm Ar}) \sim 11.4 \zeta_p(H)$.

\begin{figure}
\centering
\includegraphics[width=9cm]{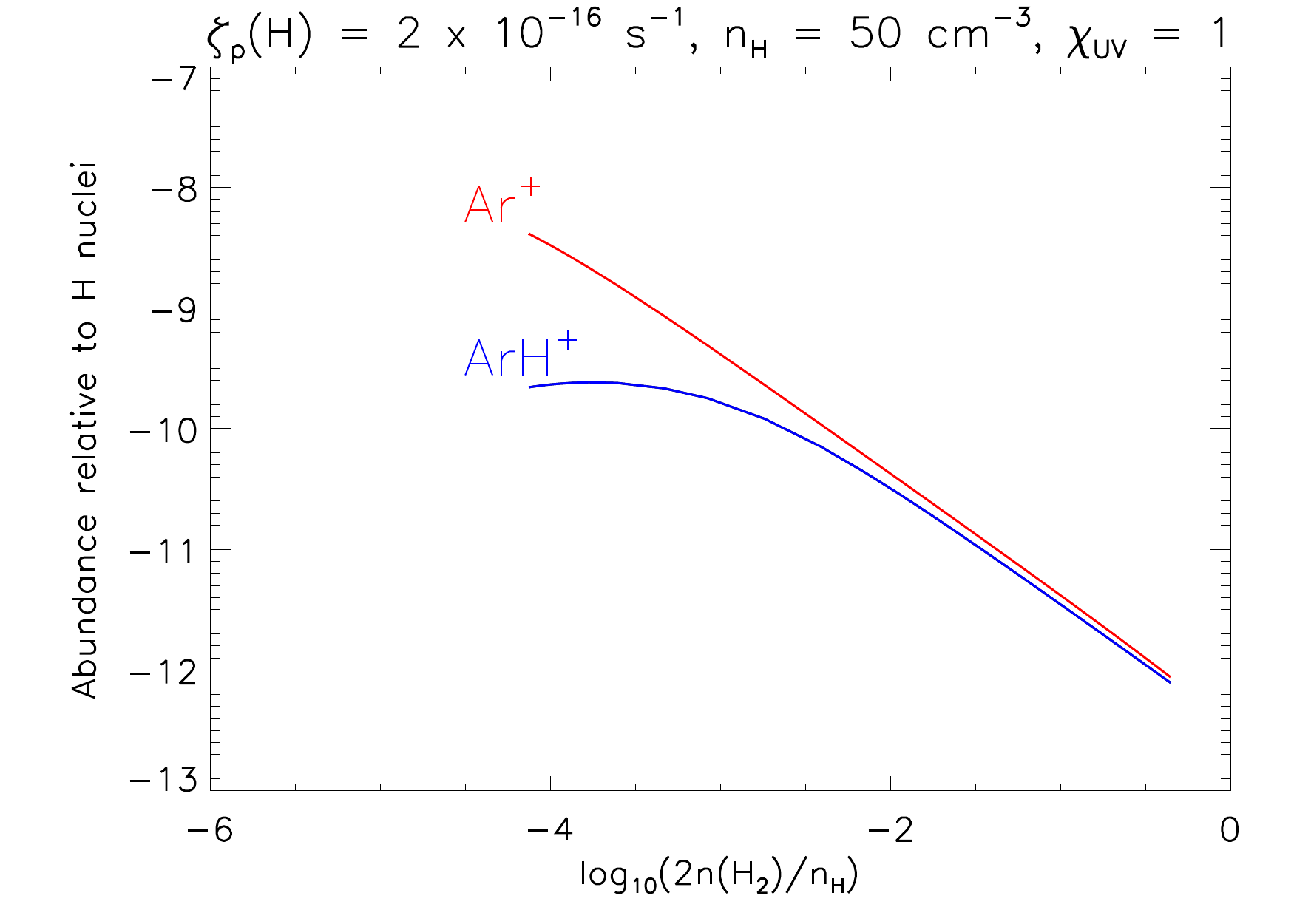}
\caption{Same as Figure~\ref{fig_a1}, except with the Ar$^+$ and ArH$^+$ abundances shown as a function of molecular fraction} \label{fig_a3}
\end{figure}

Now assuming an atomic oxygen abundance of $3.9 \times 10^{-4}$ relative to H nuclei, and assuming a solar argon abundance of $3.2 \times 10^{-6}$ (Lodders 2010), we find that equation~(\ref{eq}) may be rewritten as
\begin{equation}
 {n({\rm ArH}^+) \over n_{\rm H}} = {1.2 \times 10^{-10} \zeta_p({\rm H})_{-16} \over
n_2[1+1280 f({\rm H}_2)] + 0.35 \chi_{\rm UV} \times f_A}\label{eq2}
\end{equation}
where $n_2 = n_{\rm H} / 100\,\rm cm^{-3}$, $\zeta_p({\rm H})_{-16}=\zeta_p({\rm H})/10^{-16}\,\rm s^{-1}$, $f({\rm H}_2) = 2n({\rm H}_2)/n_{\rm H}$ is the molecular fraction, and $f_A$ is the factor by which the photodissociation rate for ArH$^+$ is reduced by attenuation.  For the isotropic illumination that we assume, $f_A=[E_2(3.6\,A_V)+E_2(3.6\,[A_V({\rm tot})-A_V])]/2$, where $E_2$ is an exponential integral; for the $A_V({\rm tot})=0.3$~mag model, $f_A$ varies from 0.56 at the cloud surface to 0.30 at the cloud center. Figure~\ref{fig_a3} shows the Ar$^+$ and ArH$^+$ abundances as a function of the molecular fraction.  For molecular fractions in excess of $\sim 10^{-4}$, equation~(\ref{eq2}) reproduces the exact behavior to within 15\%.

\begin{figure}
\centering
\includegraphics[width=9cm]{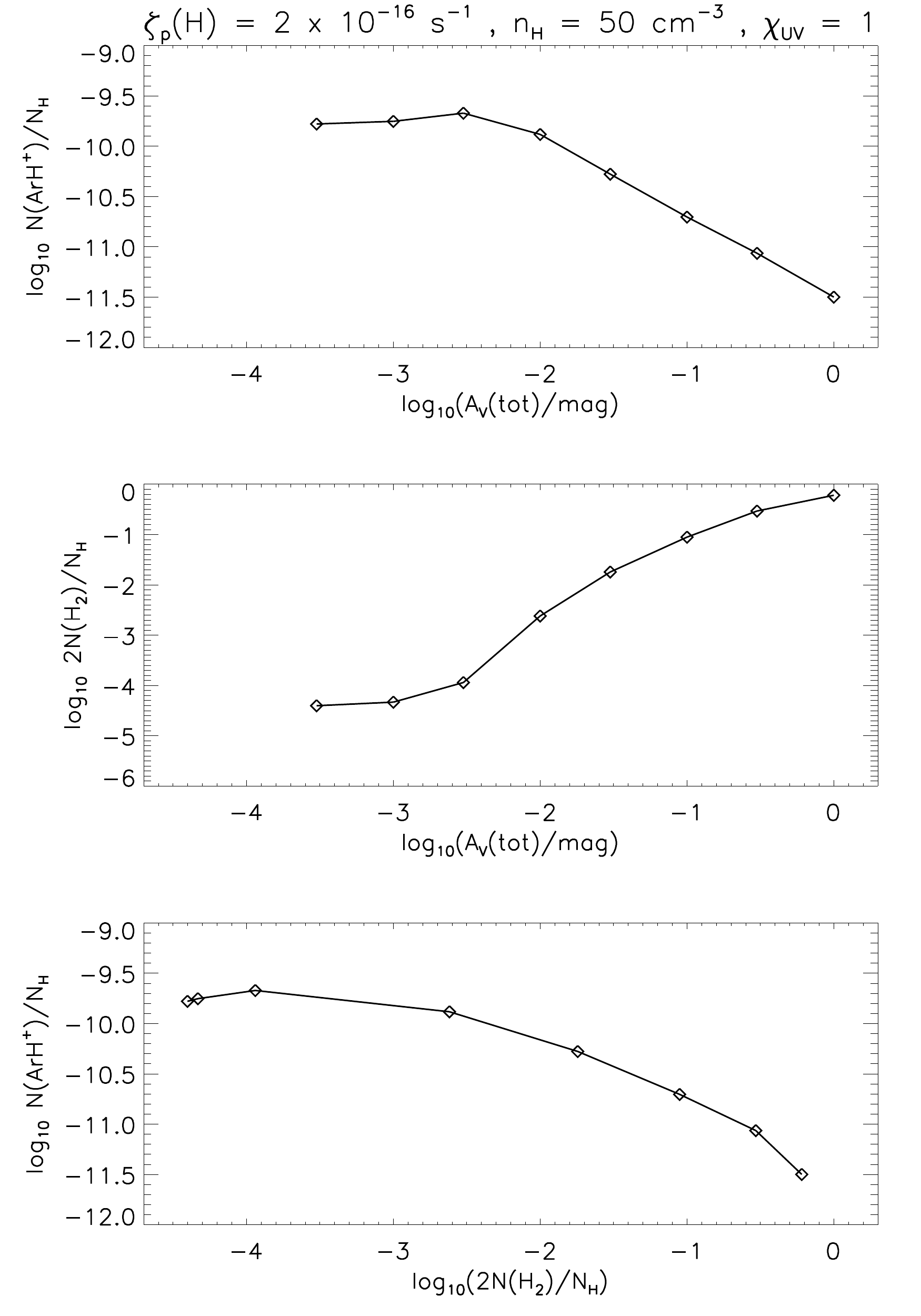}
\caption{Column-averaged ArH$^+$ abundance, $N({\rm ArH}^+)/N_{\rm H}$ (top panel), and average molecular fraction, $2N({\rm H_2})/N_{\rm H}$, as a function of $A_V({\rm tot})$.  Bottom panel: column-averaged ArH$^+$ abundance versus average molecular fraction, $2N({\rm H_2})/N_{\rm H}$  } \label{fig_a4}
\end{figure}
While astrochemical models predict molecular abundance ratios as a function of position within a gas cloud, astronomical observations measure column density ratios \new{averaged} along the line-of-sight.  Accordingly, we have calculated the column-averaged ArH$^+$ abundance, $N({\rm ArH}^+)/N_{\rm H}$, for a series of models with different assumed $A_V({\rm tot})$.  In the upper and middle panels of Figure~\ref{fig_a4}, we plot the $N({\rm ArH}^+)/N_{\rm H}$ ratio and the average molecular fraction,  $2N({\rm H_2})/N_{\rm H}$, as a function of $A_V({\rm tot})$, while in the lower panel we show the column-averaged ArH$^+$ abundance as a function of molecular fraction. 
Given the cosmic ray ionization rate $\zeta_p({\rm H}) = 2 \times 10^{-16}\,\rm s^{-1}$ and the gas density $n_{\rm H} = 50 \,\rm cm^{-3}$ assumed in our standard model,  peak $N({\rm ArH}^+)/N_{\rm H}$ ratios $\sim 2 \times 10^{-10}$ are achieved within small clouds of total visual extinction $\simlt 0.01$~mag, within which the average molecular fraction is $\simlt 10^{-3}$.  The predicted peak abundances scale linearly with the assumed cosmic ray ionization rate and, for weak UV fields, inversely with the density. For UV fields higher than $\chi_{\rm UV} = n_2[1+1200 f({\rm H}_2)]/(0.35 f_A)$, i.e., for $\chi_{\rm UV} \approx 10 n_2$, photodissociation is the dominant process.  

For the standard cosmic ray ionization rates assumed in our diffuse cloud  
models, the peak ArH$^+$ abundances, predicted to be $\approx 2 \times 10^{-10}$ relative to H,  
fall near the lower end of the observed range reported in Section 4.  Higher 
observed abundances may indicate local enhancements in the cosmic ray  
ionization rate.  Along the sight-­‐line to Sgr B2 (M),  the observed ArH$^+$/H  
ratio is largest for velocities corresponding to \textit{X2} orbits in the Galactic 
Center: these are indeed exactly the cloud velocities for which the largest  
cosmic ray ionization rate would be expected.   
 
The diffuse cloud models presented in Section~\ref{sec:chem} provide a natural  
explanation for why ArH$^+$ is present in the diffuse arm and interarm gas, but absent in the  
denser gas associated with the background continuum sources: owing to its 
rapid destruction by H$_2$, the predicted ArH$^+$ abundance falls rapidly once the 
molecular fraction exceeds $\approx  10^{-3}$.   Thus, both theory and observation  
suggest that argonium is the molecule that paradoxically abhors molecular 
clouds.  


We have also run Turbulent Dissipation Region \citep[TDR;][]{Godard2009} models, but have found that they predict no significant enhancement of ArH$^+$. This is not surprising, since the only endothermic production rate in Tab.~\ref{tab:chem}, the reaction of H$_3^+$ with Ar, is only important in regions of large molecular fraction, where ArH$^+$ is rapidly removed. Although we did not find the elevated temperatures due to enhanced  viscous dissipation and ion-neutral friction  in regions of intermittent turbulent dissipation to be important in driving endothermic reactions of relevance to the production of ArH$^+$, these regions are possible sites of cosmic ray acceleration because they are associated with intense current sheets \citep{Momferratos2014}.

\section{Summary}
We \new{confidently assign} the 617.5~GHz line to the carrier $^{36}$ArH$^+$, since
features of $^{38}$ArH$^+$ were also detected toward Sgr~B2(M) and (N) with $^{36}$Ar/$^{38}$Ar 
ratios close to, but probably smaller than in the solar neighborhood. 
The line surveys cover the frequency of the $J = 1 - 0$ transition 
of $^{20}$NeH$^+$ and even though Ne is much more abundant in space than Ar, we do not 
observe neonium absorption. This \new{difference} is in line with expectations based on the much higher ionization potential of Ne.

Our chemical calculations show that ArH$^+$ can exist only in low-density gas with a low H$_2$ fraction ($f({\rm H}_2)\approx 10^{-4}-10^{-3}$), and  a weak UV field, while an enhanced cosmic ray flux can boost its abundance.  OH$^+$ and H$_2$O$^+$ trace gas with a larger H$_2$ fraction of 0.1, and are therefore complementary probes \citep{OH_n^+_2010, Neufeld2010, OH_n+_chemistry_2012}.
 It is noteworthy, in this context, that the ArH$^+$ and H$_2$O$^+$ column densities are not well-correlated, although one would assume that ArH$^+$ and H$_2$O$^+$ both trace the stratified PDR structures of diffuse clouds, ArH$^+$ the very outer edge, and H$_2$O$^+$ gas deeper in.  It appears that this picture is too simplistic.  The aforementioned tracers  OH$^+$ and H$_2$O$^+$ trace partly molecular gas, while the so-called CO-dark gas, which is predominantly molecular, but does not contain significant abundances of CO, is best traced by HF, CH, H$_2$O, or HCO$^+$ \citep{Qin2010, Sonnentrucker2010, Flagey2013}, but also by [C{\sc ii}]  \citep{Langer2014}, which however is not very specific to this component. 
The careful analysis of column density variations in these tracers promises to \new{disentangle }the distribution of the H$_2$ fraction, providing a direct observational constraint on the poorly known transition of primarily atomic diffuse gas 
to dense molecular gas traced by CO emission,  putting strong constraints upon magnetohydrodynamical simulations for the interstellar gas \citep[e.g.][]{Micic2012, Levrier2012} and thus potentially evolving into a tool to characterize the ISM. Paradoxically, ArH$^+$ actually is a better tracer of almost purely atomic gas than the H{\sc i} line, because with the column density of H we see gas that could be 0.1\%, 1\%, or 50\% molecular, while ArH$^+$ singles out gas which is more than 99.9\% atomic.  

However, the possibilities of getting more data are limited. While both the 909~GHz OH$^+$ line and the 607~GHz \emph{para}-H$_2$O$^+$ line can be observed under very good weather conditions from very good sites on the ground \citep[see ,e.g.][]{OH+_det_2010}, ArH$^+$, due to its proximity to the 620.7~GHz water line, is extremely difficult even from excellent sites. Receivers covering these frequencies with SOFIA would therefore be highly beneficial. The other possibility to get access to these species are toward redshifted galaxies. There, however, OH$^+$ and H$_2$O$^+$ are often seen in emission, indicating very different excitation conditions. ArH$^+$ has not been found in extragalactic sources yet, but could be a very good tracer of cosmic rays in diffuse gas with little UV penetration.

\begin{acknowledgements}
   HIFI has been designed and built by a consortium of institutes and university departments from across 
Europe, Canada and the United States under the leadership of SRON Netherlands Institute for Space
Research, Groningen, The Netherlands and with major contributions from Germany, France and the US. 
Consortium members are: Canada: CSA, U.Waterloo; France: CESR, LAB, LERMA,  IRAM; Germany: 
KOSMA, MPIfR, MPS; Ireland, NUI Maynooth; Italy: ASI, IFSI-INAF, Osservatorio Astrofisico di Arcetri- 
INAF; Netherlands: SRON, TUD; Poland: CAMK, CBK; Spain: Observatorio Astron�mico Nacional (IGN), 
Centro de Astrobiolog�a (CSIC-INTA). Sweden:  Chalmers University of Technology - MC2, RSS \& GARD; 
Onsala Space Observatory; Swedish National Space Board, Stockholm University - Stockholm Observatory; 
Switzerland: ETH Zurich, FHNW; USA: Caltech, JPL, NHSC. H.S.P.M. is very grateful to the Bundesministerium f\"ur Bildung und Forschung (BMBF) for initial support through project FKZ 50OF0901 (ICC HIFI \textit{Herschel})
aimed at maintaining the Cologne Database for Molecular Spectroscopy, CDMS. This support has been administered by the Deutsches Zentrum f\"ur Luft- und Raumfahrt (DLR). Part of this work was supported by the German
   \emph{Deut\-sche For\-schungs\-ge\-mein\-schaft} in the Collaborative Research Center SFB956, and by the German Ministry of Science (BMBF) trough contract 05A11PK3. This work also has been
supported by NASA through an award issued by JPL/Caltech. We thank Christian Endres for tireless work on the molecular line catalog implementation.

\end{acknowledgements}


\end{document}